\documentclass[journal]{IEEEtran}
\usepackage{amsmath}
\usepackage{algorithm}
\usepackage{algpseudocode}
\usepackage{amssymb}
\usepackage{booktabs}
\usepackage{subfig}
\usepackage{siunitx} 
\usepackage{graphicx} 
\usepackage{subcaption}
\usepackage{algorithm}
\usepackage{float}
\usepackage{mathtools}

\begin{document}

\title{A Novel Improved Beluga Whale Optimization Algorithm for Solving Localization Problem in Swarm Robotic Systems}

\author{Zuhao Teng, Qian Dong* 
\thanks{Zuhao Teng is with the School of Advanced Technology, Xi'an Jiaotong-liverpool University, Suzhou, China.
 E-mail: Zuhao.Teng22@alumni.xjtlu.edu.cn}
\thanks{Qian Dong is with the School of Advanced Technology, Xi'an Jiaotong-liverpool University, Suzhou, China.
 E-mail: qian.dong@xjtlu.edu.cn}
\thanks{Manuscript received April 19, 2005; revised August 26, 2015.}}

\markboth{Journal of \LaTeX\ Class Files,~Vol.~14, No.~8, August~2015}%
{Shell \MakeLowercase{\textit{et al.}}: Bare Demo of IEEEtran.cls for IEEE Journals}

\maketitle

\begin{abstract}
In Swarm Robotic Systems (SRSs), only a few robots are equipped with Global Positioning System (GPS) devices, known as anchors. A challenge lies in inferring the positions of other unknown robots based on the positions of anchors. Existing solutions estimate their positions using distance measurements between unknown robots and anchors. Based on existing solutions, this study proposes a novel meta-heuristic algorithm - Improved Beluga Whale Optimization Algorithm (IBWO) to address the localization problem of SRSs, focusing on enhancing the accuracy of localization results. Simulation results demonstrate the effectiveness of this study. Specifically, we test the localization accuracy of robots under different proportions of anchors, different communication radius of robots, and different total number of robots. Compared to the traditional multilateration method and four other localization methods based on meta-heuristic algorithms, the localization accuracy of this method is consistently superior.
\end{abstract}


\IEEEpeerreviewmaketitle

\section{Introduction}

\IEEEPARstart{S}{W}arm Robotic Systems (SRSs) is a multi-robot system based on the phenomenon of swarm intelligence observed in nature \cite{duan2023animal}. This system is inspired by the collective behaviors of biological groups such as ant colonies and bird flocks. In this system, each robot operates autonomously, making decisions based on local information and interactions with neighboring robots \cite{tan2013research}. Even if individual robots fail, the overall task completion is not affected. This system has broad application prospects in fields such as search and rescue, environmental monitoring, agricultural automation, space exploration, and military reconnaissance \cite{schranz2020swarm,navarro2013introduction}. In any application, the position information of robots is essential. In fact, in most cases, robots are nearly useless without position information \cite{chen2022survey}.

SRSs are composed of hundreds of robots in general. Equipping each robot with a Global Positioning System (GPS) is impractical. Therefore, only a small subset of robots are equipped with GPS, known as anchors. The primary focus of the localization problem is on those robots not equipped with GPS, known as unknown nodes. Currently, most localization methods rely on the distances between unknown nodes and anchors. Common distance measurement methods include those based on Received Signal Strength Indicator (RSSI) \cite{mistry2015rssi} and Time of Arrival (ToA) \cite{shen2014multiple} of signals. These methods offer high distance measurement accuracy but require substantial additional hardware resources. Moreover, when there are no anchors or only a few anchors within a robot's communication radius, making single-hop measurements impossible, these distance measurement methods will fail to meet the localization requirements.

To address this issue, methods such as DV-Hop \cite{niculescu2001ad} and Sum-Dist \cite{langendoen2004distributed} have been proposed. In general, these methods approximate the distance between unknown nodes and anchor nodes by multiplying the number of hops by the average hop distance. These methods do not require additional distance measurement hardware but have lower accuracy.

Each unknown node requires the distance information from at least three anchor nodes to calculate its position. Traditional localization methods include the Multilateration \cite{wang2009localization} and the Centroid localization method \cite{zhang2012weighted}. While these methods have low computational complexity, their localization accuracy is also low. In recent years, many researchers have used meta-heuristic algorithms to solve localization problems. The main idea of these methods is to model the localization problem as a non-convex optimization problem, and by minimizing or maximizing the value of this objective function, the estimated positions of the unknown nodes can be obtained. Clearly, for these methods, the accuracy of localization largely depends on the optimization capability of the meta-heuristic algorithm.

This study proposes a novel Improved Beluga Whale Optimization Algorithm (IBWO) to address the localization problem in SRSs. The Beluga Whale Optimization Algorithm (BWO) is a meta-heuristic algorithm introduced by Zhong \textit{et al.}\cite{zhong2022beluga} in 2022. Although it possesses strong optimization capabilities, it still suffers from insufficient information sharing among search agents and a tendency to get trapped in local optimal. To address these issues, this study incorporates the Chain Foraging Strategy (CAFS) and Cyclone Foraging Strategy (CCFS) \cite{zhao2020manta} in the exploitation phase of the BWO algorithm and adds the Golden Sine Strategy (Golden-SA) \cite{tanyildizi2017golden} after the whale fall phase of BWO algorithm. The CAFS and CCFS strategies enhance information sharing among search agents, improving the algorithm's capability for detailed local search. The Golden-SA strategy increases the algorithm's ability to escape local optima. Additionally, to address the low accuracy of DV-Hop distance measurement, this study proposes the Hop-optimized DV-Hop method. In summary, the main contributions of this study are:
\begin{enumerate}
\item A novel Improved Beluga Whale Optimization Algorithm (IBWO) is proposed. Specifically, the Chain Foraging Strategy (CAFS) and the Cyclone Foraging Strategy (CCFS) are incorporated into the exploitation phase of BWO, enhancing information sharing among the population and improving the algorithm's local search capability. The Golden Sine Strategy (Golden-SA) is added after the whale fall phase of BWO, enhancing the algorithm's ability to escape local optima.
\item The Hop-optimized DV-Hop ranging method is proposed as an advancement over the traditional DV-Hop ranging method. This approach markedly enhances the precision of distance measurements.
\item A novel localization method for Swarm Robotic Systems (SRSs) is proposed. This method employs the Hop-optimized DV-Hop ranging method to measure the distance between robots and transforms the localization problem into a non-convex optimization problem. The Improved Beluga Whale Optimization Algorithm (IBWO) is then employed to optimize this problem, resulting in the estimated positions of the unknown robots.
\end{enumerate}

\section{RELATED WORK}
The importance of location information for swarm robots prompts researchers to propose effective localization methods from various perspectives. Similar to Wireless Sensor Networks (WSNs), the localization problem in SRSs requires inferring the positions of unknown robots based on the positions of anchors \cite{chen2022survey}. Therefore, the localization methods for both can largely be shared. Localization methods can be divided into two categories: range-free and range-based \cite{mao2007wireless}. The former does not involve using additional ranging hardware to directly measure distances for localization. The latter uses ranging hardware to directly measure distances to execute the localization process. The relevant research on these two types of localization methods is presented in this section.
\subsection{Range-free Localization}
Tesoriero \textit{et al.}\cite{tesoriero2010improving} propose a range-free method for indoor robots. This method is a passive RFID-based indoor localization system designed to accurately locate robots. The system places RFID tags on sensing surfaces, such as floors or walls, forming a grid where each tag represents a location unit. Robots or humans carry an RFID reader that reads these tags and sends the tag IDs to a location manager (LM) via a wireless network. The LM maps the tag IDs to physical locations and displays the positions of the entities on a virtual map. Nallanthighal \textit{et al.}\cite{nallanthighal2014improved} propose an improved grid-scan localization algorithm for wireless sensor networks. This method first collects information from 1-hop, 2-hop, and farther anchor nodes, then uses the 1-hop anchor information to estimate a region, which is subsequently divided into a grid. The valid grids are determined using information from the 1-hop and 2-hop anchors. Additionally, the information from farther anchor nodes further reduces the number of valid grids. Simulation results demonstrate that this method significantly improves localization accuracy compared to existing grid-scan algorithms \cite{yan2009grid}.

Wang \textit{et al.}\cite{wang2010research} propose a localization algorithm for wireless sensor networks based on the APIT and Monte Carlo methods. The traditional APIT algorithm suffers from low localization accuracy due to sparse anchor nodes, especially in the triangular overlap regions. To address this issue, the paper introduces the MC-APIT algorithm, which employs the Monte Carlo method for random sampling within the overlap region and filters the samples using the RSSI sequence values of the target node, ensuring that the mathematical expectation of the sample values converges to that of the target node. Simulation results indicate that the MC-APIT algorithm effectively reduces the sampling area and localization energy consumption, and significantly improves localization accuracy compared to the traditional APIT algorithm. Yuan \textit{et al.}\cite{yuan2018secure} also improve the traditional APIT algorithm. This improved algorithm enhances the APIT localization scheme by using received signal strength (RSS) to detect and defend against Sybil attacks, achieving low overhead and high detection rates. During implementation, the process begins with beacon exchange, where anchor nodes broadcast their location information and sensor nodes establish a list of neighboring anchor nodes. Next, the PIT test is conducted to determine whether the sensor node is within the anchor nodes' triangle by comparing the received signal strengths. Following this, APIT aggregation is performed, using a grid scan algorithm to combine the results of inner and outer triangles to identify the largest overlapping region. Finally, the centroid calculation is used to estimate the position of the sensor node.

The previously mentioned range-free localization methods do not utilize the distances between nodes to perform localization. In contrast, DV-Hop \cite{niculescu2001ad} use the number of hops multiplied by the average hop distance as an approximation to substitute for the distances between nodes. Given that the average hop distance does not accurately reflect the true distance, the estimation error tends to accumulate as the number of hops increases. There are already quite a few improved versions of the DV-Hop algorithm in existence. Han \textit{et al.}\cite{han2021multi} combine differential evolution and improve DV-Hop algorithms to address average distance per hop error. Then introduced mutation operations and embedded Particle Swarm Optimization for faster convergence and improved optimization. Chen \textit{et al.}\cite{chen2020cwdv} develope an iteration that modifies the average one-hop distance between anchor nodes and employs particle swarm optimization for more precise position estimation. Shi \textit{et al.}\cite{shi2018distance} improve the traditional DV-Hop scheme by using a combination of Min-max and Steepest Descent method as the node position derivation algorithm, replacing Lateration for higher localization accuracy. Cui \textit{et al.}\cite{cui2018high} presente an enhancement for IoT applications, improving hop-count values and employing Differential Evolution to address optimization challenges. Cai \textit{et al.}\cite{cai2020weight} constructe a mathematical model linking weights and hops, using a genetic algorithm for solving the model and enhancing positional precision. Due to the advantages of DV-Hop, including its simplicity, ease of implementation, and independence from additional ranging hardware, this study employs an improved DV-Hop method for distance estimation.
\subsection{Range-based Localization}
Range-based methods utilize additional ranging hardware to directly measure the distance between nodes and then estimate the position of unknown nodes based on these distances using techniques such as Multilateration \cite{wang2009localization}. Common range-based methods include received signal strength indicator (RSSI) \cite{mistry2015rssi}, time of arrival (TOA) \cite{shen2014multiple}, time difference of arrival (TDOA) \cite{sun2019wireless}, and angle of arrival (AOA) \cite{watanabe2021wireless}.

Mistry \textit{et al.}\cite{mistry2015rssi} propose an indoor localization and identification method based on Received Signal Strength Indicator (RSSI), suitable for ZigBee wireless sensor networks in smart homes. This method combines fingerprinting and proximity algorithms, determining thresholds through receiver operating characteristic analysis for room-level localization. Additionally, it estimates the actual position of the user through interaction with home system devices, and utilizes an RSSI proximity algorithm for user identification in multi-user environments. Awad \textit{et al.}\cite{awad2007adaptive} propose a distance measurement and localization method based on Received Signal Strength Indicator (RSSI). The method includes two distance estimation approaches: one based on statistical methods and the other on training artificial neural networks (ANNs). Experimental results demonstrate that both methods exhibit high accuracy and reliability in short-distance measurement and localization.

Shen \textit{et al.}\cite{shen2014multiple} address the multi-source localization problem in wireless sensor networks using Time of Arrival (TOA) measurements. The authors propose a three-step joint optimization algorithm to estimate source-measurement associations, source locations, and initial signal transmission times. The algorithm simplifies the problem through convex relaxation and approximations, achieving accuracy comparable to methods with known associations. Watanabe \textit{et al.}\cite{watanabe2021wireless} propose a novel localization method for wireless sensor networks based on angle-of-arrival (AoA) measurements. This method first uses the least squares technique for initial position estimation, then improves localization accuracy by weighting each anchor and evaluation function term based on the calculated error variance. Sun \textit{et al.}\cite{sun2019wireless} propose a wireless sensor network localization method based on Time Difference of Arrival (TDOA) measurements. The authors introduced a modified polar representation (MPR) method, achieving high-precision localization using Maximum Likelihood Estimation (MLE) and a two-step Weighted Total Least Squares (WTLS) estimator. This method does not require prior knowledge of whether the signal source is in the near-field or far-field.

\section{The basis of proposed localization method}
\subsection{DV-Hop Method}
The DV-Hop method is a renowned range-free method, which principally unfolds in three phases:

\begin{enumerate}
\item Each anchor node disseminates data packets containing its location information using flooding techniques, with an initial hop count of 1. Upon receiving such data packets, each unknown node logs the hop count, the anchor node's location and the minimum hop count determined thus far. If a newly received data packet has a hop count smaller than the previously recorded hop count, the unknown node records this smaller hop count and continues to forward the data packet, incrementing the hop count by 1. If the received hop count is greater than the recorded hop count, the data packet is discarded. This method ensures that each unknown node in the network obtains the minimum hop count to each anchor node.

\item Each anchor node calculates an average hop size, as shown in Eq.(1).
\begin{equation}
\text{Hopesize}_i = \frac{\sum \sqrt{(x_{i}-x_{j})^2 - (y_{i}-y_{j})^2}}{\sum h_{ij}} \quad (i \neq j).
\end{equation}
Where \(i, j\) denote distinct anchors, the coordinates \((x_{i}, y_{i})\) and \((x_{j}, y_{j})\) denote the positions of these anchors, while \(h_{ij}\) represents the hop counts between them. Following the estimation of this inferred hop length, the average hop size is subsequently distributed to all unknown nodes across the network.

\item Each unknown node calculates its distance to the anchor nodes based on the minimum hop count and the average hop size, as shown in Eq.(2).
\begin{equation}
d_i = K_i \times \text{Hopesize}_i
\end{equation}
Where \( K_i \) denotes the hop count between unknown node \( i \) and anchors, \( d_i \) represents the distance between unknown node \( i \) and anchors, and \( \text{Hopesize}_i \) signifies the average distance per hop obtained in Eq.(1).
\end{enumerate}

\subsection{Beluga Whale Optimization}
Inspired by the collective hunting and migration processes of beluga whales, Zhong \textit{et al.}\cite{zhong2022beluga} propose the Beluga Whale Optimization Algorithm (BWO) in 2022. It comprises three phases: exploration, exploitation, and whale fall, each crafted to emulate the natural behaviors of Beluga whales. In the mathematical framework of the BWO, the positions of the Beluga whales are depicted by search agents initially placed at random positions, expressed in Eq.(3).
\begin{equation}
X_{i,j} = Lb_j + (Ub_j - Lb_j) \times \text{rand}
\end{equation}
Where \( X_{i,j} \) represents the \( i \)-th beluga whale in the \( j \)-th dimension. \( Lb_j \) and \( Ub_j \) denote the lower and upper limits of the designated domain. \textit{rand} denotes a random value within the range of (0,1).

The optimization process of any metaheuristic algorithm involves iterations, where the fitness values of search agents are compared, leading to continuous updates of their optimal positions. In the BWO, the fitness value matrix of search agents can be expressed in Eq.(4).
\begin{equation}
F_x = \begin{pmatrix}
f(x_{1,1}, x_{1,2}, \ldots, x_{1,d}) \\
f(x_{2,1}, x_{2,2}, \ldots, x_{2,d}) \\
\vdots \\
f(x_{n,1}, x_{n,2}, \ldots, x_{n,d})
\end{pmatrix}
\end{equation}
Where \( f(x_{1,1}, x_{1,2}, \ldots, x_{1,d}), \ldots, f(x_{n,1}, x_{n,2}, \ldots, x_{n,d}) \) represent the fitness value for each search agents.

The exploration phase of BWO is inspired by the behavior of Beluga whales swimming in pairs. The iterative process of this phase can be represented in Eq.(5).
{
\begin{equation}
\left\{
    \begin{array}{ll}
        X^{t+1}_{i,j} = X^{t}_{i,pj} + (X^{t}_{r,p1} - X^{t}_{i,pj}) (1 + r_1) \sin(2\pi r_2), &  j =\text{ even} \\
        X^{t+1}_{i,j} = X^{t}_{i,pj} + (X^{t}_{r,p1} - X^{t}_{i,pj}) (1 + r_1) \cos(2\pi r_2), &  j =\text{ odd}
    \end{array}
\right.
\end{equation}
}
Where \( X_{i,j}^{t+1} \) represents the updated position of the \( i \)-th beluga whale in \( j \)-th dimension, \( X_{i,p_j}^{t} \) denotes the current location of the \( i \)-th beluga whale. The variables \( r_1 \) and \( r_2 \) denote randomly generated numbers within the range of 0 to 1. \( X_{r,p_1}^{t} \) indicates a beluga whale selected randomly.

The position update procedure during the exploitation phase of BWO draws inspiration from the collective hunting behavior of Beluga whales. Meanwhile, within this phase, the Levy Flight strategy is adopted to enhance the probability of successful hunting endeavors among the whales. This phase can be expressed in Eq.(6).
\begin{equation}
X_{i}^{t+1} = r_3 \cdot X_{best}^{t} - r_4 \cdot X_{i}^{t} + C_1 \cdot L_F \cdot (X_{r}^{t} - X_{i}^{t})
\end{equation}
Where \( X_{best}^{t} \) represents the best position currently discovered by the beluga whales, \( X_{i}^{t} \) denotes the current position of the \( i \)-th bulage whale, and \( r_{3} \) and \( r_{4} \) are random values ranging between 0 and 1. \( X_{r}^{t} \) signifies a search agent selected randomly. The parameter \( C_{1} \) is a dynamically adjustable coefficient, and its calculation is defined in Eq.(7).
\begin{equation}
C_{1} = 2r_{4}\left(1 - \frac{t}{t_{max}}\right)
\end{equation}
Where \( t \) is the current iteration count and \( t_{max} \) is the maximum number of iterations allowed.

\( L_F \) refers to the Levy flight function, which can be computed in Eq.(8).
\begin{equation}
L_F = 0.05 \times \frac{u \times \sigma}{|v|^{1/\beta}}
\end{equation}

\begin{equation}
\sigma = \left( \frac{\Gamma(1+\beta) \times \sin(\pi \beta / 2)}{\Gamma((1+\beta)/2) \times \beta \times 2^{(\beta-1)/2}} \right)^{1/\beta}
\end{equation}
Where \( B_0 \) is a random number generated between 0 and 1 during each iteration, and \( t \) is the count of the current iteration out of a total of \( t_{\text{max}} \), which is the maximum iterations set. If the factor \( B_f > 0.5 \), the search agents 
will adjust their positions for exploration phase. Conversely, if \( B_f < 0.5 \), the exploitation phase is applied.

The Whale Fall phase refers to the event when a deceased beluga whale sinks to the ocean's depths. This can occur as beluga whales face threats from polar bears, or  human activities during their feeding and migratory journeys. When this happens, the whales' carcasses settle on the sea floor. The new positions of the beluga whales during this phase are depicted in Eq.(10).
\begin{equation}
X_{i}^{t+1} = r_5 X_{i}^t - r_6 X_{r}^t + r_7 X_{\text{step}}
\end{equation}
Where \( r_5 \), \( r_6 \), and \( r_7 \) stand for random values selected between 0 and 1. The variable \( X_{\text{step}} \) indicates the magnitude of the search agents' descent, which can be calculated in Eq.(11).
\begin{equation}
X_{\text{step}} = (u_b - l_b) \exp\left(-C_2 \frac{T}{T_{\text{max}}}\right)
\end{equation}
Where \( u_b \) and \( l_b \) represent the upper and lower bounds of the control variables, 
respectively. \( C_2 \) is a dynamic factor influenced by the size of the population and the 
likelihood of whale fall occurrences, and it can be expressed in Eq.(12).

\begin{equation}
C_2 = 2W_f \times n
\end{equation}
\begin{equation}
W_f = 0.1 - 0.05t / t_{\text{max}}
\end{equation}
Where \( n \) refers to the size of the population, \( t \) indicates the present iteration cycle, 
and \( t_{\text{max}} \) is the total number of iterations permitted.

The likelihood of a whale fall diminishes progressively from 0.1 at the start of the iterations to 0.05 by the end, 
suggesting that as search agents approach the food source during the optimization process, their risk decreases.

\section{Proposed localization method}
\subsection{Hop-optimized DV-Hop}
In the traditional DV-Hop method, the distance estimation between nodes is based on the hop counts and average hop size. However, due to the uneven distribution of nodes in SRSs, the estimation of average hop size often contains significant errors, which accumulate as the hop count increases. These errors directly affect the accuracy of node localization. In complex environments, the errors can be further amplified. To enhance the accuracy of node localization in SRSs, it is necessary to improve the hop count calculation rule to more accurately estimate the distances between nodes.

In Hop-optimized DV-Hop, the hop count between anchors and other nodes within anchors' communication radius is redefined, as shown in Eq.(14).
\begin{equation}
h_{o,M} =
\begin{cases} 
\frac{1}{3}, & M(x_{M}, y_{M}) \subseteq \theta \\
\frac{2}{3}, & M(x_{M}, y_{M}) \subseteq \delta \\
1, & M(x_{M}, y_{M}) \subseteq \gamma
\end{cases}
\end{equation}
Where \( h_{O,M} \) represents the hop count between anchor O and node M, which is located within the communication radius of O. $(x_{M}, y_{M})$ are the coordinates of node M. $\theta$
, $\delta$, and $\gamma$ are three regions within the communication radius of anchor O, as defined in Eq.(15).
\begin{equation}
\begin{cases}
\theta = \{(x, y) \mid RSSI(x, y) < RSSI_{\frac{R}{3}}\} \\
\delta = \{(x, y) \mid RSSI_{\frac{R}{3}} \leq RSSI(x, y) < RSSI_{\frac{2R}{3}}\} \\
\gamma = \{(x, y) \mid RSSI_{\frac{2R}{3}} \leq RSSI(x, y) < RSSI_{R}\}
\end{cases}
\end{equation}
Where \((x, y)\) represent a set of points. \(RSSI(x, y)\) represents the received signal strength indicator at point \((x, y)\) from anchor O. \(RSSI_{\frac{R}{3}}, RSSI_{\frac{2R}{3}},\) and \(RSSI_{R}\) represent the received signal strength at distances \(\frac{R}{3}, \frac{2R}{3},\) and \(R\) from anchor O, respectively.

\begin{figure}[!htb]
  \centering
  \includegraphics[width=1\linewidth]{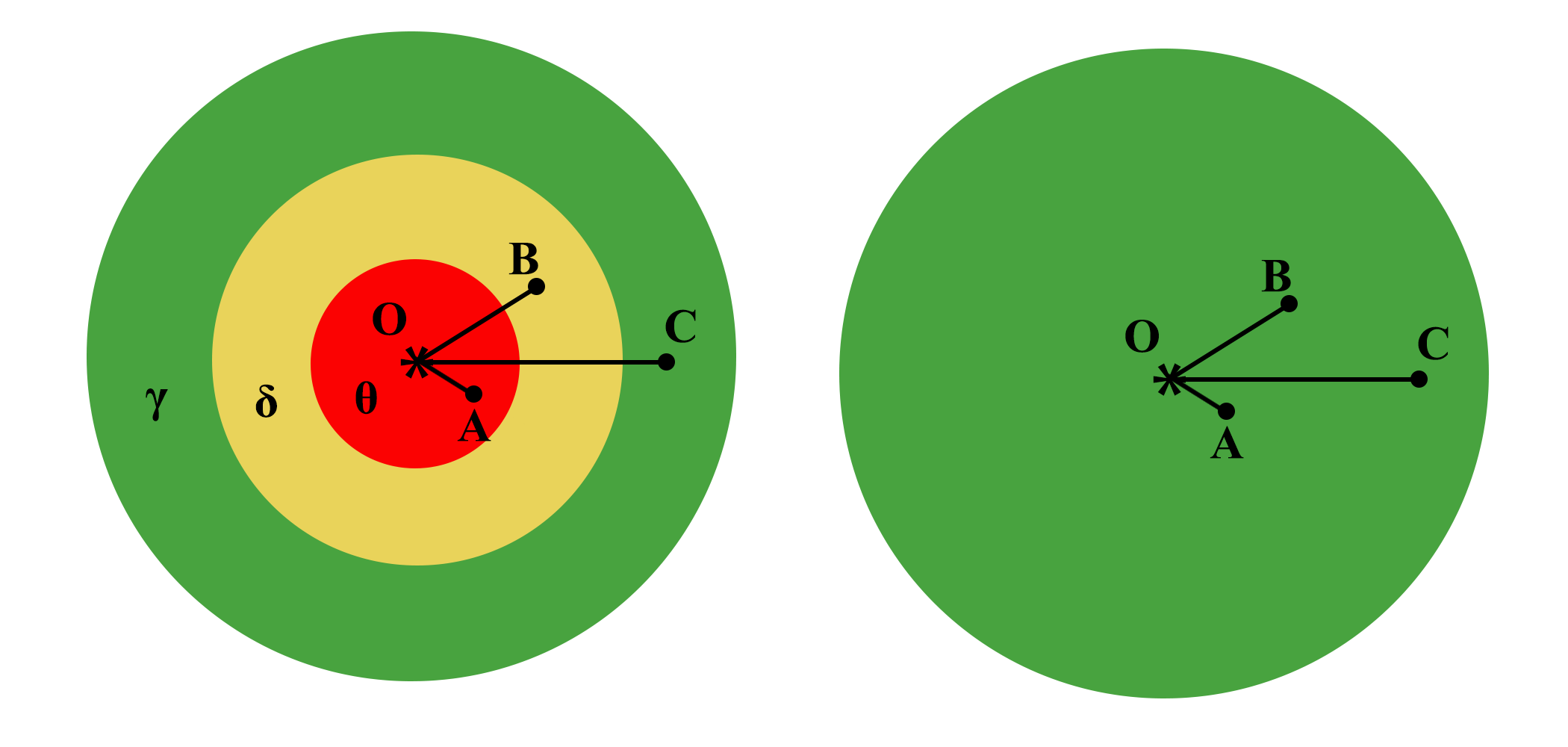}
  \caption{(a) Hop-optimized DV-Hop. (b) DV-Hop.}
  \label{hop_optimized}
\end{figure}

$\theta$, $\delta$, $\gamma$ are segmented based on the distribution of signal strength from anchor nodes, with signal strength being approximately equal at the same concentric circle \cite{dong2023interior}. Firstly, the RSSI is measured on circular rings with radii of R/3 and 2R/3 to serve as thresholds for segmenting the region. Other nodes within the communication radius of the anchor determine their hop count to the anchor by identifying which RSSI threshold range they fall into. As shown in Fig.\ref{hop_optimized}(a), the hop counts from nodes A, B, and C to anchor O are 1/3, 2/3, and 1, respectively. In contrast, in the original DV-Hop method, their hop counts to the anchor O are all 1, as shown in Fig.\ref{hop_optimized}(b).

It can be seen that the original single hop is refined into more precise fractions. Since this type of distance measurement method uses the hop count multiplied by the average hop distance as the distance estimate, a more accurate hop count can yield more precise distance estimation results.

\subsection{Improved Beluga Whale Optimization Algorithm}
The Improved Beluga Whale Optimization Algorithm (IBWO) is developed on the integration of three optimization strategies: the Cyclone Foraging Strategy (CCFS), the Chain Foraging Strategy (CAFS) and the Golden Sine Strategy (Golden-SA).
\subsubsection{Cyclone Foraging Strategy}
\begin{figure}[!htb]
  \centering
  \includegraphics[width=0.75\linewidth]{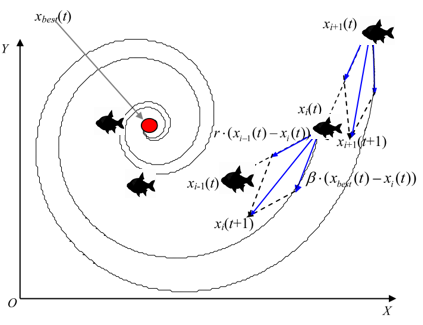} 
  \caption{CCFS strategy}
  \label{CCFS}
\end{figure}

The Cyclone Foraging Strategy (CCFS) draws its design from the spiraling patterns observed in group behaviors near optimal solutions found in numerous optimization algorithms \cite{zhao2020manta}. The application of CCFS enhances the exploitation capabilities of the traditional BWO algorithm by inducing a spiral search behavior in the search agents. This behavior not only revolves around the optimal solution but also progresses in a directed manner towards the adjacent leading search agent. Fig.\ref{CCFS} depicts the CCFS in action, illustrating the search agents' spiral paths. And it can be modeled as follows:

{
\footnotesize
\begin{equation}
X^{t+1}_i = 
\begin{cases}
X^{t}_{\text{best}} + r_8 (X^{t}_{\text{best}} - X^{t}_i) + \beta (X^{t}_{\text{best}} - X^{t}_i), & \text{for } i = 1 \\
X^{t}_{\text{best}} + r_8 (X_{i-1}^{t} - X^{t}_i) + \beta (X^{t}_{\text{best}} - X^{t}_i), & \text{for } i = 2, \ldots, n
\end{cases}
\end{equation}

\begin{equation}
\beta = 2e^{r_9 \frac{t_{\text{max}}-t+1}{t_{\text{max}}}} \cdot \sin(2\pi r_9)
\end{equation}
}
where \( \beta \) serves as a scale factor, while \( r_8 \) and \( r_9 \) are random variables within the [0,1] range. For the \( (t+1) \)-th iteration cycle, \( X^{t+1}_i \) specifies the \( i \)-th search agent's position. The optimal position of the search agent during the \( t \)-th iteration is denoted by \( X^{t}_{\text{best}} \), where \( X^{t}_i \) and \( X^{t}_{i-1} \) represent the positions of the \( i \)-th and the \( (i-1) \)-th search agents, respectively, at the \( t \)-th iteration. The symbol \( t_{\text{max}} \) indicates the maximum number of allowable iteration cycles.

\subsubsection{Chain Foraging Strategy}
\begin{figure}[H]
  \centering
  \includegraphics[width=0.75\linewidth]{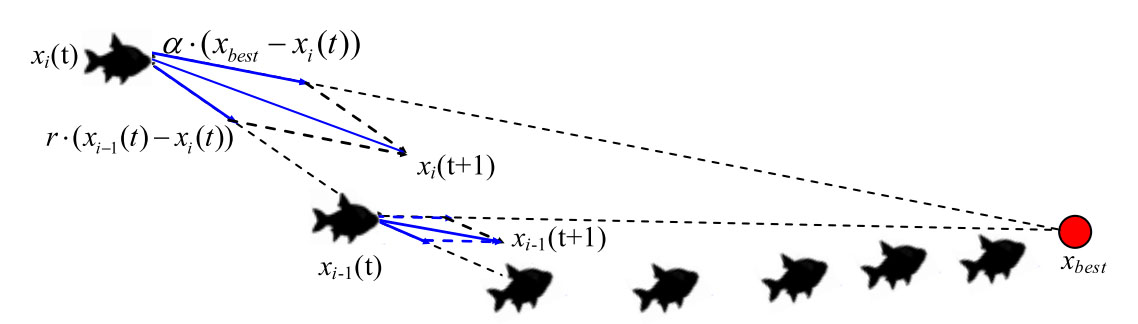} 
  \caption{CAFS strategy}
  \label{CAFS}
\end{figure}

The Chain Foraging Strategy (CAFS) is an optimization technique inspired by the collective foraging behavior of animals in nature \cite{zhao2020manta}. This strategy involves arranging each search agent in a line, one after the other. Each search agents navigates not only towards the direction of the current optimal solution but also moves in the direction of the search agents ahead in the sequence, as illustrated in Fig.\ref{CAFS}. Similar to the CCFS, the CAFS contributes to the enhancement of the exploitation phase within optimization algorithms. The CAFS is mathematically modeled as follows:

{
\footnotesize
\begin{equation}
X^{t+1}_i = 
\begin{cases}
X^{t}_{\text{i}} + r_9 (X^{t}_{\text{best}} - X^{t}_i) + \alpha (X^{t}_{\text{best}} - X^{t}_i), & \text{for } i = 1 \\
X^{t}_{\text{i}} + r_9 (X_{i-1}^{t} - X^{t}_i) + \alpha (X^{t}_{\text{best}} - X^{t}_i), & \text{for } i = 2, \ldots, n
\end{cases}
\end{equation}
\begin{equation}
\alpha = 2 \cdot r_9 \cdot \sqrt{\left| \log(r_9) \right|}
\end{equation}
}
where \( X^{t}_i \) represents the location of the \( i \)-th search agent at iteration \( t \), and \( r_9 \) is a uniformly distributed random vector within the [0, 1] range. The term \( \alpha\) acts as a weighting factor, and \( X^{t}_{\text{best}} \) indicates the position of the best solution in t-th iteration.

\subsubsection{Golden Sine Strategy}
The Golden Sine Algorithm (Golden-SA) \cite{tanyildizi2017golden} is a mathematically inspired search strategy. This strategy is inspired by the golden ratio and the sine function, incorporating the golden ratio coefficient and the sine function into the position update process. This integration introduces greater randomness and nonlinear variability into the search process, thereby enhancing the algorithm's ability to escape local optima. The position update formula for the Golden Sine Algorithm is shown in Eq.(22).
\begin{equation}
X_i^{t+1} = X_i^t \left| \sin(r_1) \right| - r_2 \sin(r_1) \left| x_1 X_{\text{best}}^t - x_2 X_i^t \right|
\end{equation}
where $X_i^{t+1}$ represents the position of the $i$-th search agent at the $(t+1)$-th iteration. $X_i^t$ denotes the position of the $i$-th search agent at the $t$-th iteration. $X_{\text{best}}^t$ refers to the position of the best search agent up to the $t$-th iteration. $r_1$ is a random number within the range $[0, 2\pi]$, and $r_2$ is a random number within the range $[0, \pi]$. $x_1$ and $x_2$ are golden ratio coefficients, calculated as shown in Eq.(23) and Eq.(24).
\begin{equation}
x_1 = a \cdot (1 - t) + b \cdot t
\end{equation}
\begin{equation}
x_2 = a \cdot t + b \cdot (1 - t)
\end{equation}
where $a = -\pi$ and $b = \pi$. $t$ is the golden ratio, defined as $t = \frac{\sqrt{5} - 1}{2}$.

\begin{table*}[!htbp]
\caption{Characteristics of 23 classical benchmark functions}
\label{benchmark}
\renewcommand{\arraystretch}{1.5}
\begin{tabular}{@{}lllll@{}}
\toprule
Function & Function equation                               & Dim & Range        & Optimal \\ \midrule
Unimodal & & & & \\
F1       & \( f_1(x) = \sum_{i=1}^{n}x_i^2 \)              & 30  & \([-100,100]\) & 0       \\
F2       & \( f_2(x) = \sum_{i=1}^{n}|x_i| + \prod_{i=1}^{n}|x_i| \) & 30  & \([-10,10]\)    & 0       \\
F3       & \( f_3(x) = \sum_{i=1}^{n}\left(\sum_{j=1}^{i}x_j\right)^2 \) & 30  & \([-100,100]\) & 0       \\
F4       & \( f_4(x) = \max |x_i|, \quad 1 \leq i \leq n \) & 30  & \([-100,100]\) & 0       \\
F5       & \( f_5(x) = \sum_{i=1}^{n-1} \left[ 100(x_{i+1} - x_i^2)^2 + (x_i - 1)^2 \right] \) & 30  & \([-30,30]\) & 0       \\
F6       & \( f_6(x) = \sum_{i=1}^{n} (x_i + 0.5)^2 \) & 30  & \([-100,100]\) & 0       \\
F7       & \( f_7(x) = \sum_{i=1}^{n} i x_i^4 + \text{random}(0, 1) \) & 30  & \([-1.28,1.28]\) & 0       \\
Multimodal & & & & \\
F8       & \( f_8(x) = \sum_{i=1}^{n} -x_i \sin(\sqrt{|x_i|}) \) & 30  & \([-500,500]\)  & \( -418.9829 \times 5 \) \\
F9       & \( f_9(x) = \sum_{i=1}^{n} \left[x_i^2 - 10\cos(2\pi x_i) + 10\right] \) & 30  & \([-5.12,5.12]\)  & \( 0 \) \\
F10       & \( f_{10}(x) = -20\exp\left(-0.2\sqrt{\frac{1}{n}\sum_{i=1}^{n}x^i}\right) - \exp\left(\frac{1}{n}\sum_{i=1}^{n}\cos(2\pi x_i)\right) + 20 + e \) & 30  & \([-32,32]\)  & \( 0 \) \\
F11       & \( f_{11}(x) = \frac{1}{4000}\sum_{i=1}^{n}x_i^2 - \prod_{i=1}^{n}\cos\left(\frac{x_i}{\sqrt{i}}\right) + 1 \) & 30  & \([-600,600]\)  & \( 0 \) \\
F12       & \(\begin{aligned}[t]
f_{12}(x) &= \frac{\pi}{n}\biggl\{10\sin(\pi y_1) + \sum_{i=1}^{n-1}\left[(y_i - 1)^2\left(1 + 10\sin^2(\pi y_{i+1})\right)\right] + (y_n - 1)^2\biggr\} \\
&\quad + \sum_{i=1}^{n}\mu(x_i, 10, 100, 4) \\
&\quad y_i = 1 + \frac{x_i + 1}{4} \\
&\quad \mu(x_i, a, k, m) = 
\begin{cases} 
k(x_i - a)^m & x_i > a \\
0 & -a \leq x_i \leq a \\
k(-x_i - a)^m & x_i < -a 
\end{cases}
\end{aligned}\) & 30 & [-50,50] & 0 \\
F13       & \( f_13(x) = \sum_{i=1}^{n} -x_i \sin(\sqrt{|x_i|}) \) & 30  & \([-50,50]\)  & \( 0 \) \\
Fixed-dimension multimodal & & & & \\
F14      & \( f_{14}(x) = \frac{1}{500} + \sum_{j=1}^{25}\left(\frac{1}{j + \sum_{i=1}^{2}(x_i - a_{ij})^6}\right) - \frac{1}{\sum_{j=1}^{25}\frac{1}{j}} \) & 2   & \([-65,65]\)    & 1       \\
F15      & \( f_{15}(x) = \sum_{i=1}^{11} \left[ a_i - \frac{x_1(b_i^2 + b_ix_2)}{b_i^2 + b_ix_3 + x_4} \right]^2 \) & 4   & \([-5,5]\)      & 0.00030 \\
F16      & \( f_{16}(x) = 4x_1^2 - 2.1x_1^4 + \frac{1}{3}x_1^6 + x_1x_2 - 4x_2^2 + 4x_2^4
 \) & 2   & \([-5,5]\)      & -1.0316 \\
F17      & \( f_{17}(x) = \left( x_2 - \frac{5.1}{4\pi^2}x_1^2 + \frac{5}{\pi}x_1 - 6 \right)^2 + 10 \left( 1 - \frac{1}{8\pi} \right)\cos x_1 + 10 \) & 2   & \([-5,5]\)      & 0.398 \\
F18      & \(
f_{18}(x) = \left[1 + (x_1 + x_2 + 1)^2(19 - 14x_1 + 3x_1^2 - 14x_2 + 6x_1x_2 + 3x_2^2)\right] 
\) & 2   & \([-2,2]\)      & 3 \\

      & \( \times \left[30 + (2x_1 - 3x_2)^2(18 - 32x_1 + 12x_1^2 + 48x_2 - 36x_1x_2 + 27x_2^2)\right]
 \)  \\

F19      & \( f_{19}(x) = - \sum_{i=1}^{4} c_i \exp \left( - \sum_{j=1}^{3} a_{ij}(x_i - p_{ij})^2 \right)
 \) & 3   & \([-1,2]\)      & -3.86 \\
F20      & \( f_{20}(x) = - \sum_{i=1}^{4} c_i \exp \left( - \sum_{j=1}^{6} a_{ij}(x_j - p_{ij})^2 \right)
\)      & 6 & \([0,1]\)      & -3.32\\
F21      & \( f_{21}(x) = - \sum_{i=1}^{5} \left[ (X - a_i)(X - a_i)^T + c_i \right]^{-1}
\)      & 4 & \([0,10]\)      & -10.1532\\
F22      & \( f_{22}(x) = - \sum_{i=1}^{7} \left[ (X - a_i)(X - a_i)^T + c_i \right]^{-1}
\)      & 4 & \([0,10]\)      & -10.4028\\
F23      & \( f_{23}(x) = - \sum_{i=1}^{10} \left[ (X - a_i)(X - a_i)^T + c_i \right]^{-1}
\)      & 4 & \([0,10]\)      & -10.5363\\
\bottomrule
\end{tabular}
\end{table*}

\subsubsection{Improved Beluga Whale Optimization}
Despite the unique features of each meta-heuristic algorithm, the majority share a common characteristic: the search process is divided into exploration and exploitation phases \cite{abdel2018metaheuristic}. The first phase seeks to encourage important and random changes in the candidate solutions, enhance their diversity, and a thorough search of the space. This means that they should focus on the exploration of every corner in the search area in order not to congregate too early around one part of the local space. Exploitation then continues to raise the quality of solutions by searching more in the neighborhood of the promising options found during exploration.

On the other hand, the exploration phase and the exploitation phase also have distinct features. Excess exploration will allow the search agents to explore larger parts of the search area but might have poor local search capability, which in turn could result in a poor-quality final solution. Also, the excess exploitation phase can lead to the search agents possessing very narrow search ranges, thereby not guaranteeing the quality of the solution.

Meta-heuristic algorithms are generally applied to solve non-convex optimization problems, which means the objective function has multiple extreme values. Therefore, the ability of the algorithm to perform fine local searches and escape local optima is crucial. In other words, the position updates during the exploitation phase have a significant impact on the optimization capability for non-convex problems. Considering the limited information sharing among search agents during the exploitation phase of BWO, we incorporate the CCFS and CAFS strategies in this phase. These strategies fully account for information sharing among search agents, enhancing the algorithm's local search capability. Additionally, to improve BWO's ability to escape local optima, we add the Golden-SA strategy following the whale fall phase of BWO. Algorithm 1 shows the pseudocode of the IBWO algorithm.

\begin{figure*}[!t]
    \centering
    \subfloat[F1]{\includegraphics[width=1.65in]{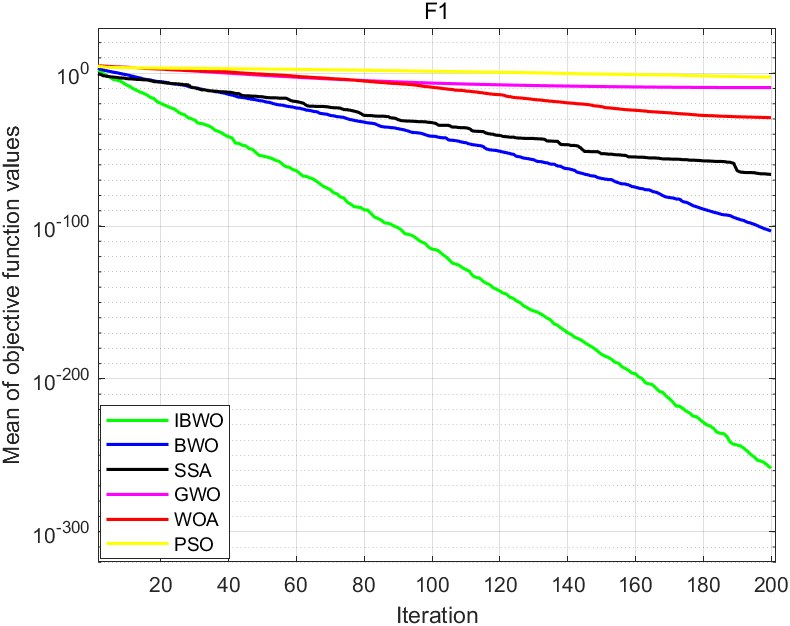}}%
    \label{figure1}
    \hfil
    \subfloat[F2]{\includegraphics[width=1.65in]{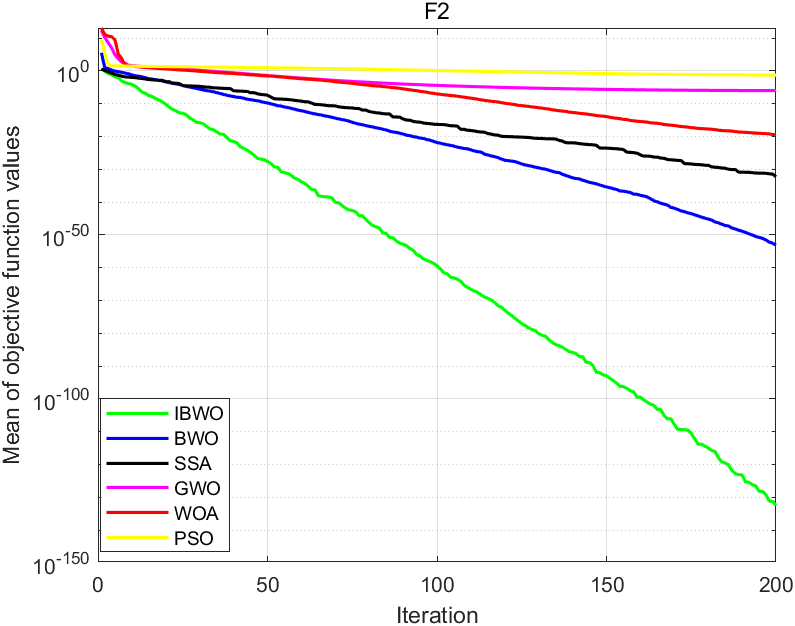}}%
    \label{figure2}
    \hfil
    \subfloat[F3]{\includegraphics[width=1.65in]{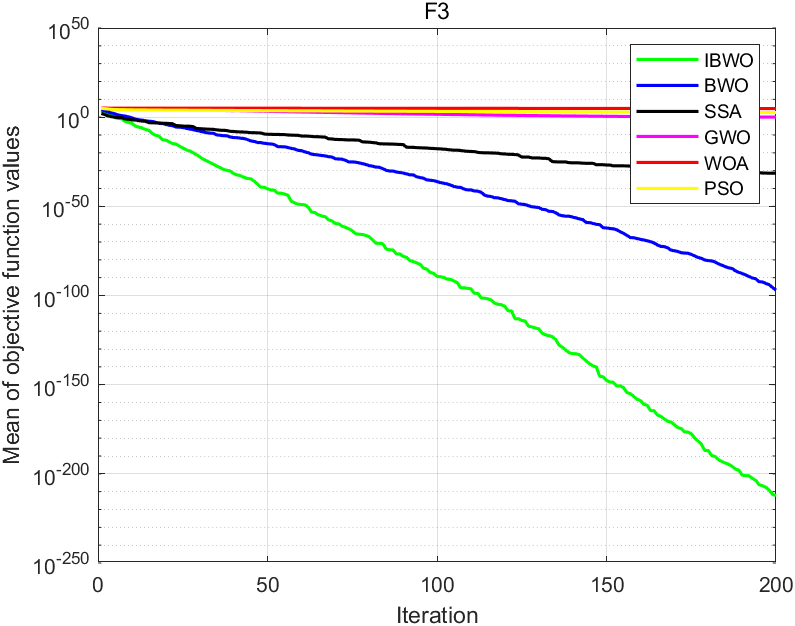}}%
    \label{figure3}
    \hfil
    \subfloat[F4]{\includegraphics[width=1.65in]{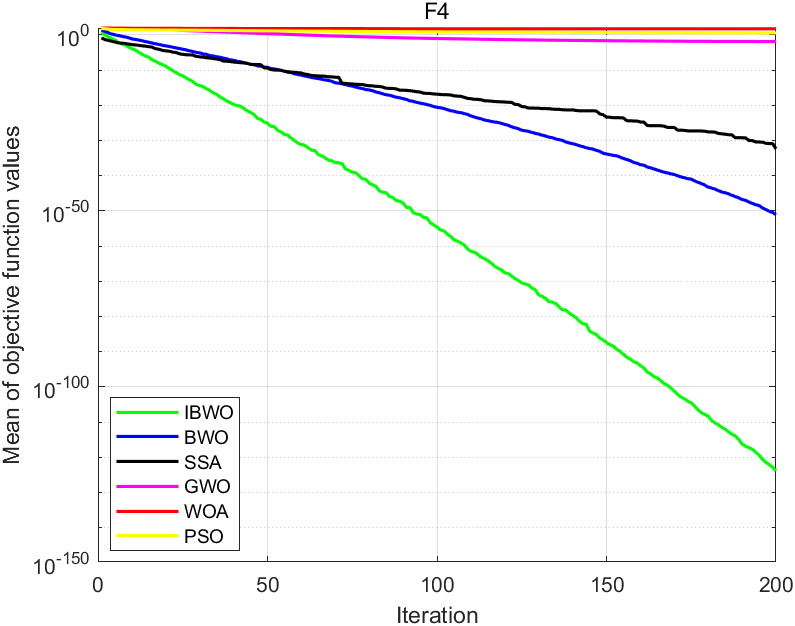}}%
    \label{figure4}\\
    \subfloat[F5]{\includegraphics[width=1.65in]{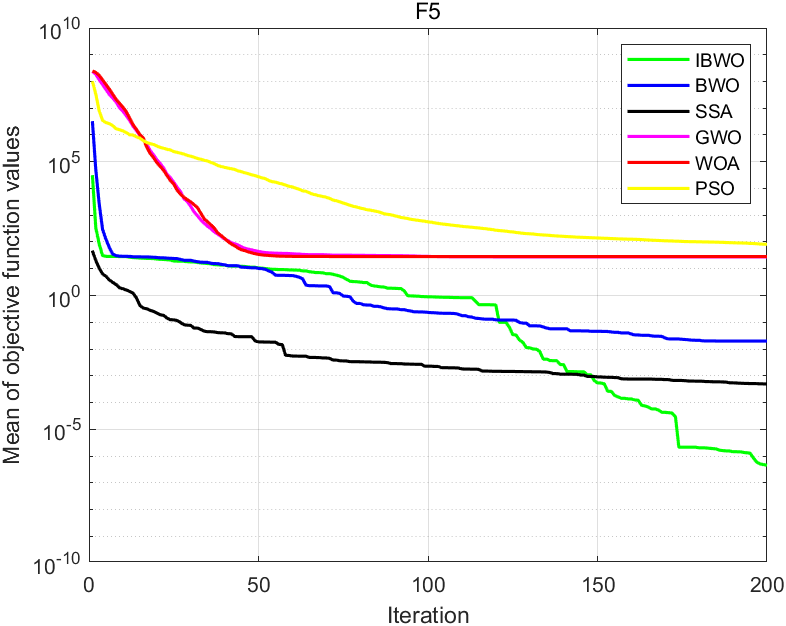}}%
    \label{figure5}
    \hfil
    \subfloat[F6]{\includegraphics[width=1.65in]{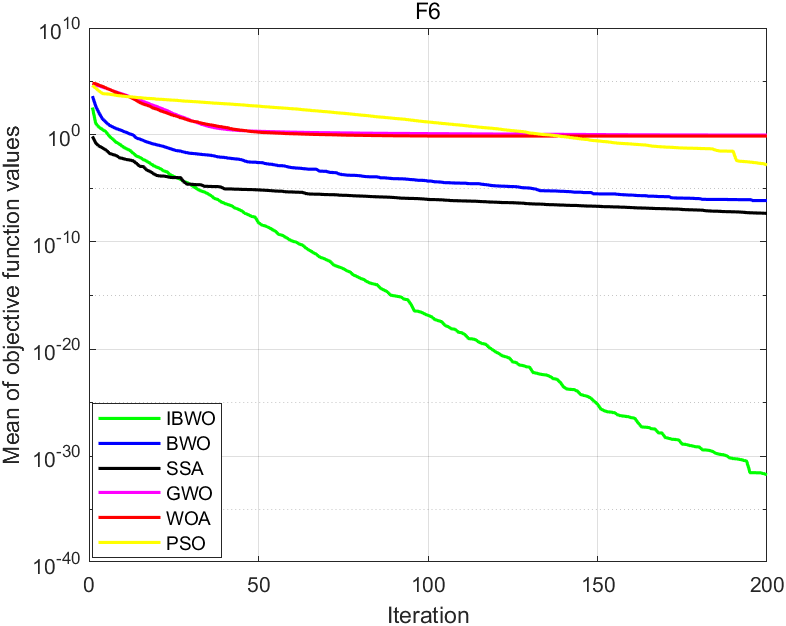}}%
    \label{figure6}
    \hfil
    \subfloat[F7]{\includegraphics[width=1.65in]{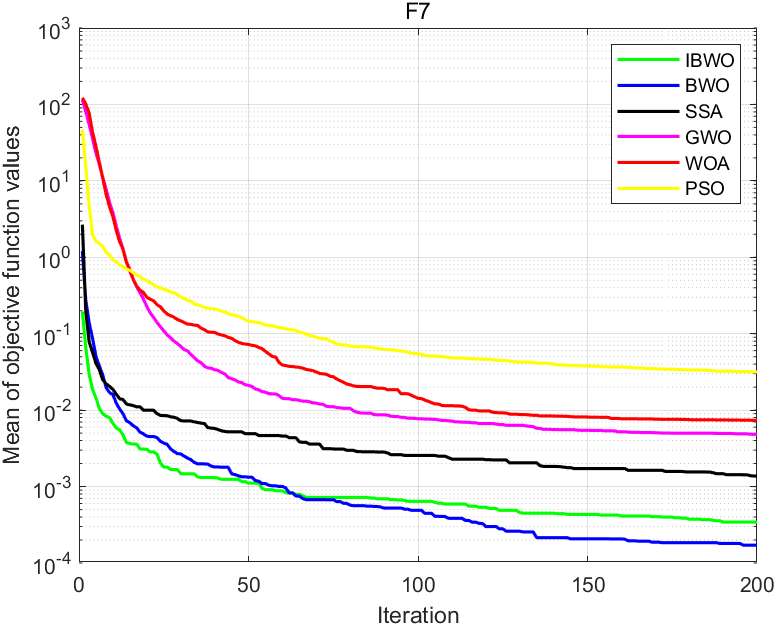}}%
    \label{figure7}
    \hfil
    \subfloat[F8]{\includegraphics[width=1.65in]{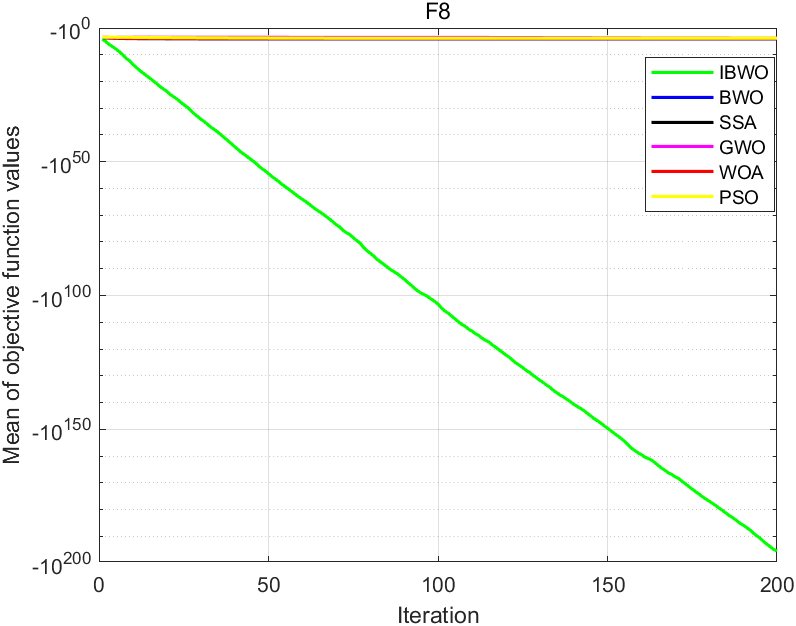}}%
    \label{figure8}\\
    \subfloat[F9]{\includegraphics[width=1.65in]{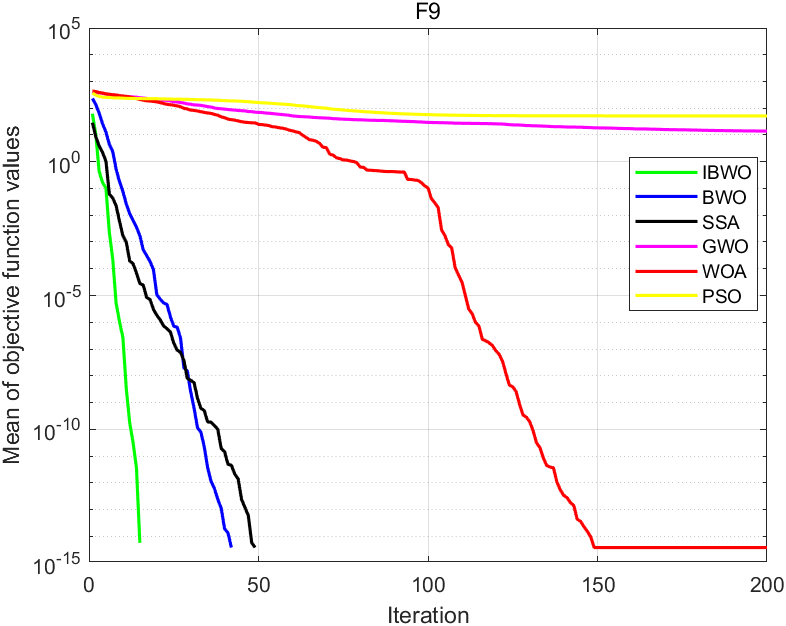}}%
    \label{figure9}
    \hfil
    \subfloat[F10]{\includegraphics[width=1.65in]{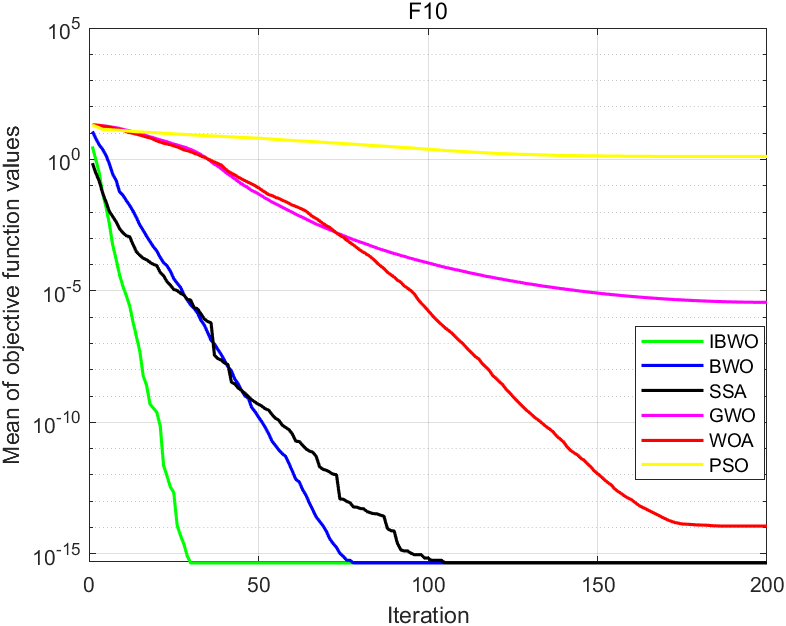}}%
    \label{figure10}
    \hfil
    \subfloat[F11]{\includegraphics[width=1.65in]{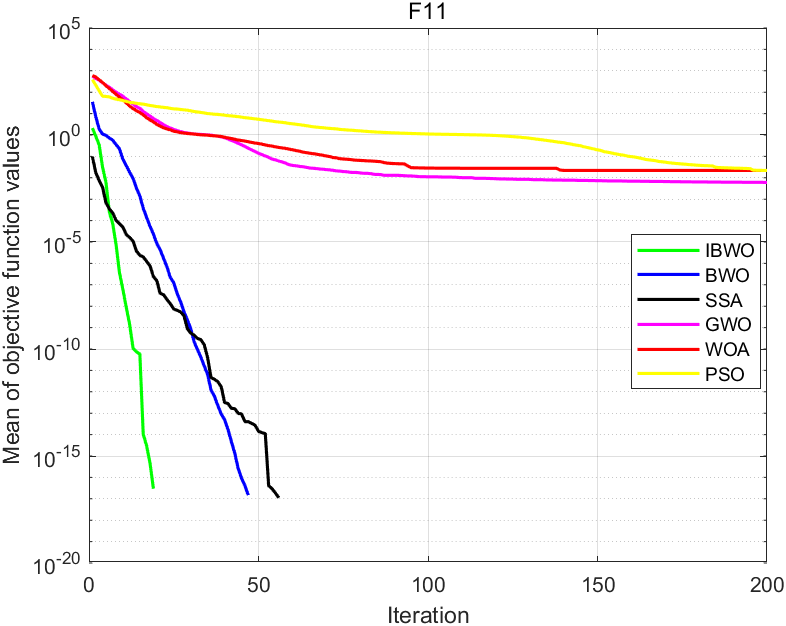}}%
    \label{figure11}
    \hfil
    \subfloat[F12]{\includegraphics[width=1.65in]{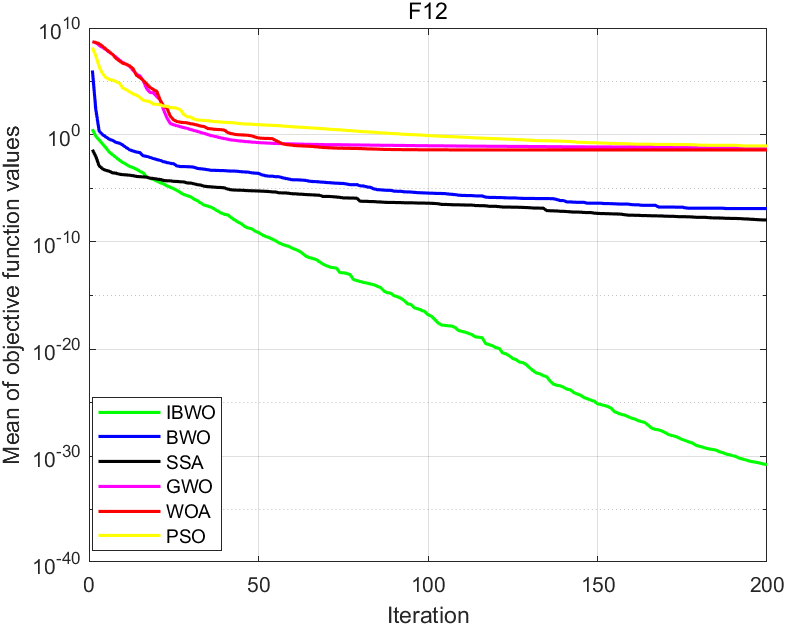}}%
    \label{figure12}\\
    \subfloat[F13]{\includegraphics[width=1.65in]{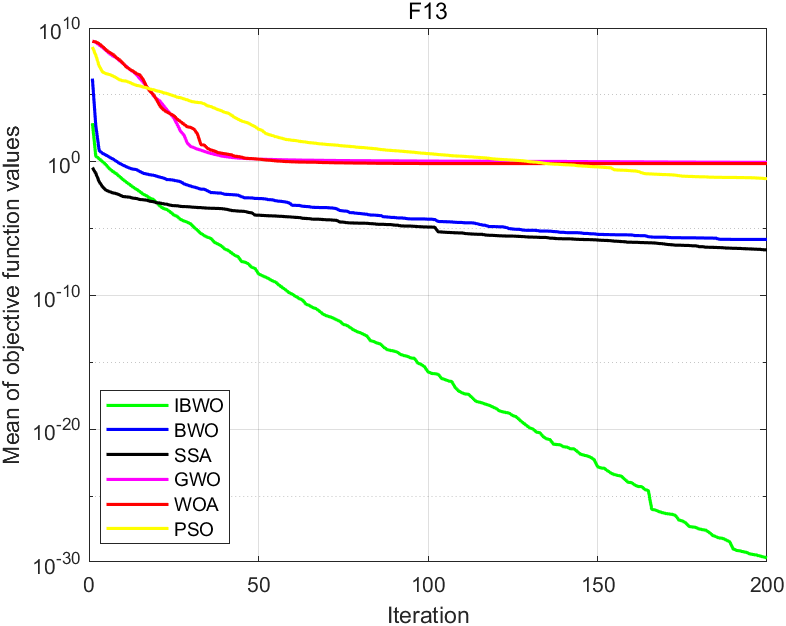}}%
    \label{figure13}
    \hfil
    \subfloat[F14]{\includegraphics[width=1.65in]{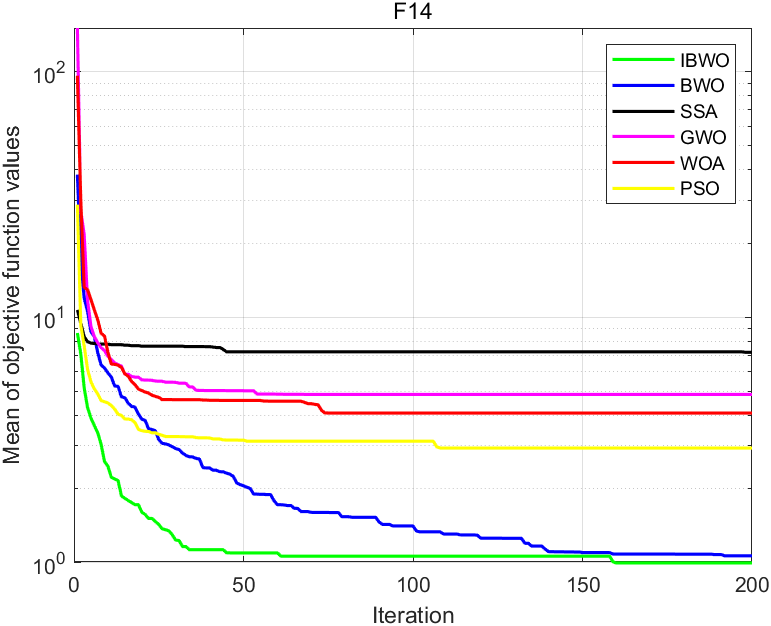}}%
    \label{figure14}
    \hfil
    \subfloat[F15]{\includegraphics[width=1.65in]{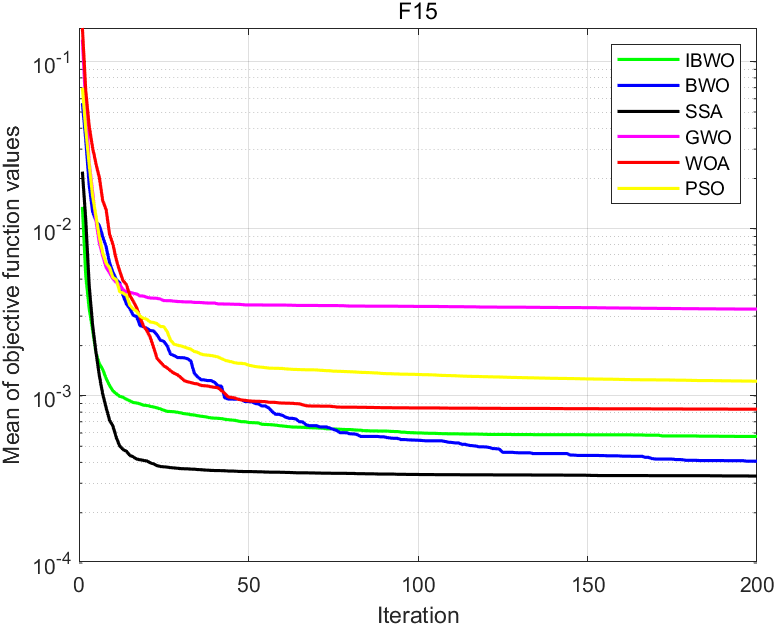}}%
    \label{figure15}
    \hfil
    \subfloat[F16]{\includegraphics[width=1.65in]{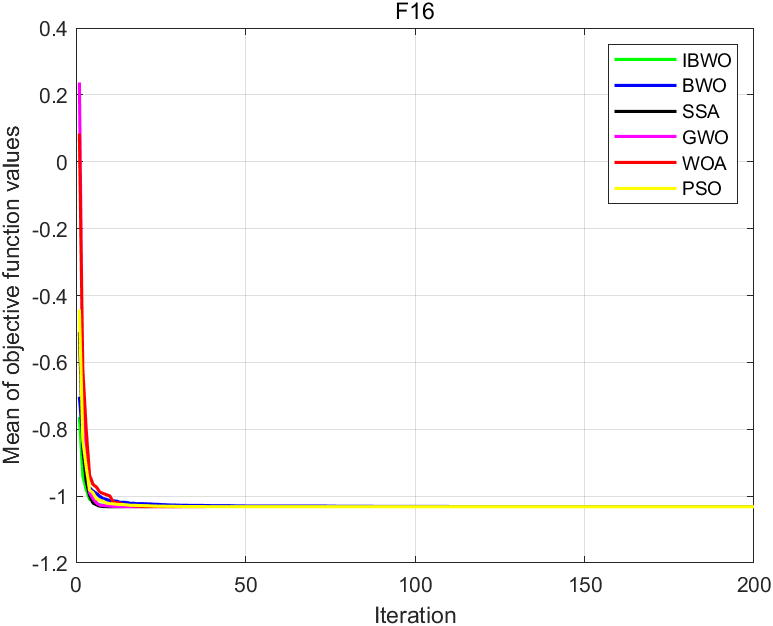}}%
    \label{figure16}\\
    \subfloat[F17]{\includegraphics[width=1.65in]{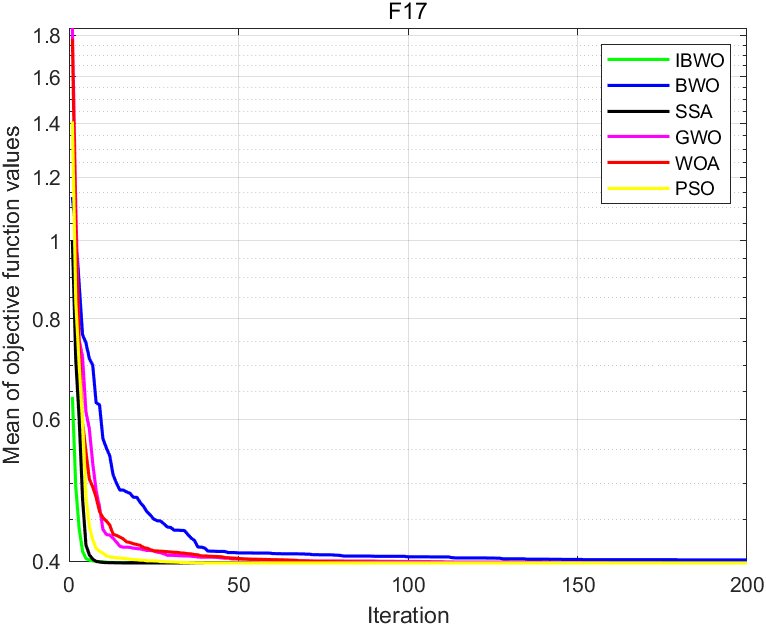}}%
    \label{figure17}
    \hfil
    \subfloat[F18]{\includegraphics[width=1.65in]{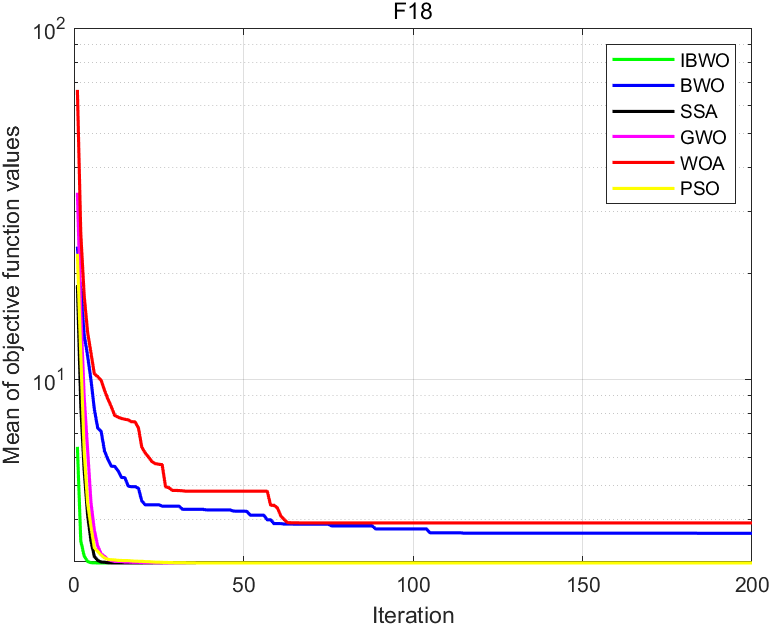}}%
    \label{figure18}
    \hfil
    \subfloat[F19]{\includegraphics[width=1.65in]{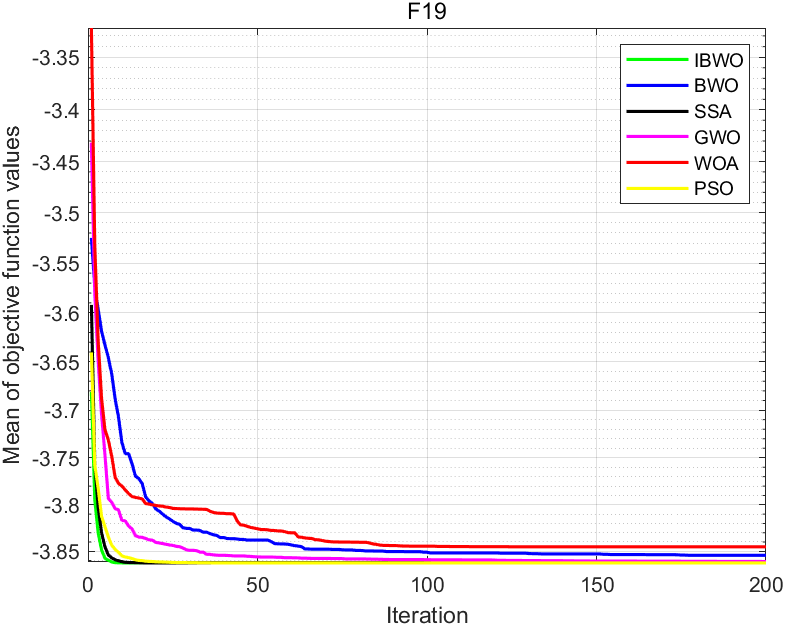}}%
    \label{figure19}
    \hfil
    \subfloat[F20]{\includegraphics[width=1.65in]{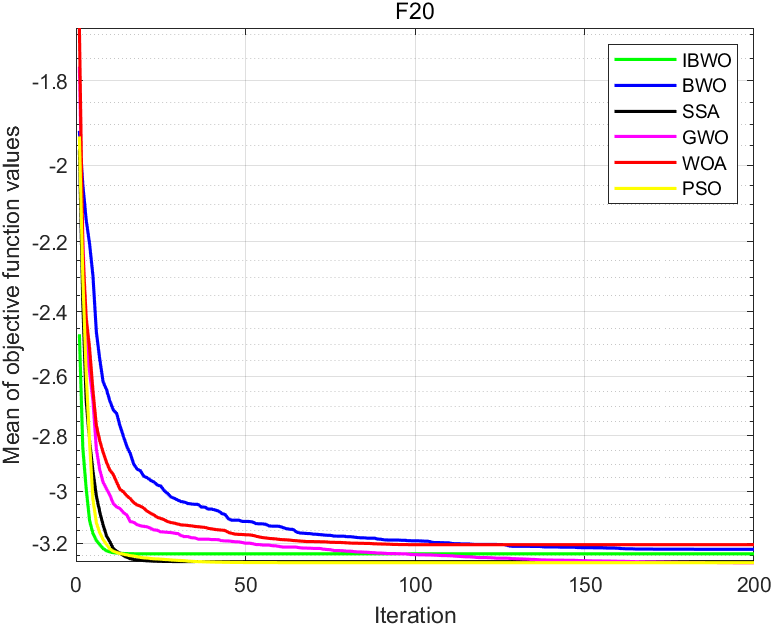}}%
    \label{figure20}\\
    \caption{The performance of IBWO and 5 other representative meta-heuristic algorithms on the F1-F20 benchmark functions.}
    \label{IBWO_graphics1}
\end{figure*}

\begin{algorithm}
\caption{Pseudocode of the IBWO algorithm}
\begin{algorithmic}[1]
\State \textbf{Input:} Algorithmic parameters
\State \textbf{Output:} The best solution
\State Initiate the population, assess the fitness values.
\While{\(T \leq T_{\max}\)}
\State Obtain \(W_f\), \(B_f\)
\For{each search agents \(X_i\)}
\If{\(B_f(i) > 0.5\)}
\State //Exploration phase
\State Generate \(p\) randomly from dimension
\State Choose a beluga whale \(X_r\) randomly
\State Update new positions using Eq. (5)
\Else
\State //Exploitation phase
\State Update \(C_1\), calculate the Levy flight function
\State Update new positions using Eq. (6)
\State //Cyclone Foraging Strategy
\State Update new positions using Eq. (16)
\State //Chain Foraging Strategy
\State Update new positions using Eq. (18)
\EndIf
\State Check the boundaries, calculate the fitness values
\EndFor
\For{each beluga whale \(X_i\)}
\State // Whale fall phase
\If{\(B_f(i) \leq W_f\)}
\State Update the step factor \(C_2\)
\State Calculate the step size \(X_{\text{step}}\)
\State Update new positions using Eq. (10)
\State Check the boundaries, calculate fitness value
\EndIf
\State // Golden Sine Strategy
\State Update new positions using Eq. (20)
\EndFor
\State Find the current best solution
\State \(T = T+1\)
\EndWhile
\State Output the best solution
\end{algorithmic}
\end{algorithm}

\subsection{Fitness Function and Localization Process}
We model the swarm robotics localization problem as a non-convex optimization problem, where the position of each unknown robot needs to be independently determined. During the optimization algorithm's iterative process, each search agent updates its position by updating its fitness value, ultimately using the position of the search agent with the lowest fitness value as the position of the unknown robot. The objective function for this localization method is shown in Eq.(25).
\begin{equation}
f(x, y) = \min \left( \sum_{j=1}^{M} \left| \sqrt{(x - x_j)^2 + (y - y_j)^2} - d_j \right| \right)
\end{equation}
where the term \( f(x, y) \) denotes the objective function associated with the search agent, which is defined by the coordinate \( (x, y) \). The coordinates \( (x_j, y_j) \) represent the location of the \( j \)-th anchor. \( d_j \) represents the distance between the j-th anchor and the unknown robot. And \( d_j \) is determined using Hop-optimized DV-Hop. \( M \) stands for the total count of anchors.

And the fitness function can be expressed as:
\begin{equation}
fitness = f(x, y)
\end{equation}

The flow chart of the Improved Beluga Whale Localization Method (IBWOL) proposed in this study is outlined in Fig.7.
\begin{figure}[!htb]
  \centering
  \includegraphics[width=\linewidth]{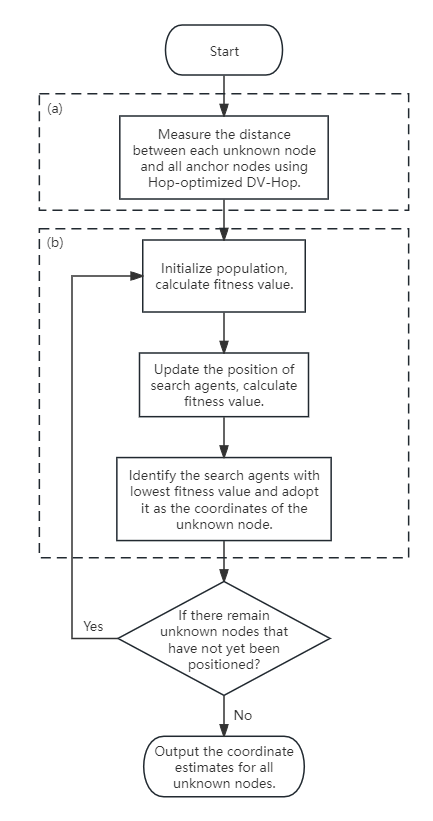}
  \caption{Flow chart of proposed localization method  (a)Hop-optimized DV-Hop  (b)Improved Beluga Whale Optimization}
  \label{fig:anchor}
\end{figure}

\begin{table*}[!htbp]
\centering
\caption{The performance results based on 23 classical benchmark functions}
\label{IBWO_table}
\begin{tabular}{@{}lcccccccc@{}}
\toprule
 & \multicolumn{2}{c}{F1} & \multicolumn{2}{c}{F2} & \multicolumn{2}{c}{F3} & \multicolumn{2}{c}{F4} \\ 
 & Avg & Std & Avg & Std & Avg & Std & Avg & Std \\ 
\midrule
IBWO & 3.58e-259 & 0 & 3.30e-133 & 1.78e-132 & 5.62e-213 & 0 & 1.66e-124 & 4.12e-124 \\
BWO & 8.55e-105 & 4.07e-104 & 6.30e-54 & 2.19e-53 & 4.09e-98 & 1.83e-97 & 4.16e-52 & 7.39e-52 \\
PSO & 1.86e-3 & 2.03e-3 & 5.48e-2 & 7.92e-2 & 4.48e+2 & 3.30e+2 & 4.07 & 1.39 \\
GWO & 3.04e-10 & 3.14e-10 & 1.12e-06 & 4.62e-07 & 1.58 & 2.55 & 1.56e-2 & 7.72e-3 \\
SSA & 5.80e-67 & 2.27e-66
 & 5.64e-33 & 2.98e-32 & 3.94e-32 & 2.16e-31 & 4.23e-33 & 2.31e-32 \\
WOA & 1.09e-29 & 5.19e-29 & 1.01e-20 & 3.85e-20 & 7.34e+4 & 1.87e+4 & 5.41e1 & 2.36e1 \\
\midrule
 & \multicolumn{2}{c}{F5} & \multicolumn{2}{c}{F6} & \multicolumn{2}{c}{F7} & \multicolumn{2}{c}{F8} \\ 
 & Avg & Std & Avg & Std & Avg & Std & Avg & Std \\ 
\midrule
IBWO & 4.48e-07 & 2.45e-06 & 1.76e-32 & 9.04e-32 & 6.95e-4 & 7.25e-4 & -2.35e+195 & 1.26e+128 \\
BWO & 1.04e-2 & 1.20e-2 & 9.80e-7 & 1.24e-6 & 2.43e-4 & 1.90e-4 & -1.26e+4 & 2.07e-1 \\
PSO & 8.36e+1 & 9.30e+1 & 6.70e-3 & 2.56e-2 & 3.43e-2 & 1.55e-2 & -6.61e+3 & 5.53e+2 \\
GWO & 2.77e+1 & 8.39e-1 & 9.84e-1 & 2.79e-1 & 4.70e-3 & 2.40e-3 & -5.56e+3 & 1.09e+3 \\
SSA & 8.10e-4 & 1.38e-3 & 4.72e-08 & 6.70e-08 & 1.79e-3 & 1.26e-3 & -9.78e+3 & 1.76e+2 \\
WOA & 1.09e-29 & 5.19e-29 & 7.84e-1 & 2.52e-1 & 9.68e-3 & 1.15e-2 & -1.00e+4 & 1.71e+3 \\
\midrule
 & \multicolumn{2}{c}{F9} & \multicolumn{2}{c}{F10} & \multicolumn{2}{c}{F11} & \multicolumn{2}{c}{F12} \\ 
 & Avg & Std & Avg & Std & Avg & Std & Avg & Std \\ 
\midrule
IBWO & 0 & 0 & 4.44e-16 & 0 & 0 & 0 & 9.57e-31 & 1.04e-31 \\
BWO & 0 & 0 & 4.44e-16 & 0 & 0 & 0 & 1.46e-07 & 2.01e-07 \\
PSO & 5.15e+1 & 1.43e+1 & 1.12 & 0.92 & 2.85e-2 & 2.90e-2 & 1.41e-1 & 2.97e-1 \\
GWO & 1.29e+1 & 8.63 & 3.86e-6 & 1.97e-6 & 9.02e-3 & 1.77e-2 & 8.71e-2 & 9.82e-2 \\
SSA & 0 & 0 & 4.44e-16 & 0 & 0 & 0 & 2.00e-8 & 2.50e-8 \\
WOA & 9.47e-15 & 3.02e-14 & 1.52e-14 & 9.19e-15 & 2.20e-2 & 8.50e-2 & 5.15e-2 & 4.29e-2 \\
\midrule
 & \multicolumn{2}{c}{F13} & \multicolumn{2}{c}{F14} & \multicolumn{2}{c}{F15} & \multicolumn{2}{c}{F16} \\ 
 & Avg & Std & Avg & Std & Avg & Std & Avg & Std \\ 
\midrule
IBWO & 1.22e-30 & 3.33e-30 & 9.98e-1 & 4.98e-17 & 4.89e-4 & 2.99e-4 & -1.03 & 4.70e-16 \\
BWO & 1.07e-6 & 1.06e-6 & 1.29 & 4.47e-1 & 4.09e-4 & 6.62e-5 & -1.03 & 4.86e-4 \\
PSO & 1.23e-1 & 2.67e-1 & 3.74 & 3.86 & 1.16e-3 & 3.64e-3 & -1.03 & 6.52e-16 \\
GWO & 8.85e-1 & 3.04e-1 & 4.47 & 3.85 & 6.50e-3 & 9.23e-3 & -1.03 & 9.23e-08 \\
SSA & 4.33e-7 & 6.56e-7 & 7.20 & 5.30 & 3.20e-4 & 7.10e-5 & -1.03 & 5.68e-16 \\
WOA & 6.77e-1 & 2.13e-1 & 4.20 & 3.96 & 1.84e-3 & 4.07e-3 & -1.03 & 4.86e-4 \\
\midrule
 & \multicolumn{2}{c}{F17} & \multicolumn{2}{c}{F18} & \multicolumn{2}{c}{F19} & \multicolumn{2}{c}{F20} \\ 
 & Avg & Std & Avg & Std & Avg & Std & Avg & Std \\ 
\midrule
IBWO & 3.98e-1 & 0 & 3 & 6.69e-16 & -3.86 & 2.26e-15 & -3.24 & 5.79e-2 \\
BWO & 4.02e-1 & 4.70e-3 & 3.83 & 6.51e-1 & -3.85 & 6.24e-3 & -3.24 & 4.88e-2 \\
PSO & 3.98e-1 & 0 & 3 & 1.88e-15 & -3.86 & 2.64e-15 & -3.27 & 5.92e-2 \\
GWO & 3.99e-1 & 5.11e-3 & 3 & 1.15e-4 & -3.86 & 2.85e-3 & -3.27 & 6.75e-2 \\
SSA & 3.98e-1 & 0 & 3 & 2.17e-15 & -3.86 & 2.48e-15 & -3.28 & 5.83e-2 \\
WOA & 3.98e-1 & 5.19e-29 & 3.90 & 4.94 & -3.85 & 3.31e-2 & -3.23 & 1.00e-1 \\
\midrule
 & \multicolumn{2}{c}{F21} & \multicolumn{2}{c}{F22} & \multicolumn{2}{c}{F23} \\ 
 & Avg & Std & Avg & Std & Avg & Std \\ 
\midrule
IBWO & -10.15 & 3.21e-15 & -10.40 & 2.86e-16 & -10.53 & 2.24e-15  \\
BWO & -9.47 & 8.50e-1 & -9.79 & 6.60e-1 & -9.78 & 9.30e-1  \\
PSO & -4.66 & 3.37 & -8.04 & 3.44 & -6.87 & 3.83  \\
GWO & -8.89 & 2.61 & -1.04e+1 & 2.86e-9 & -10.52 & 4.91e-3  \\
SSA & -7.94 & 2.44 & -7.74 & 2.14 & -9.09 & 2.65  \\
WOA & -8.26 & 2.54 & -7.12 & 3.14 & -6.48 & 3.19  \\
\bottomrule
\end{tabular}
\end{table*}

\section{Simulation Results}
This section will primarily delve into the optimization performance of the IBWO algorithm and the performance of the localization algorithm proposed in this study. All experiments in this study are conducted on a computer equipped with an Intel(R) Core(TM) i5-10300H CPU operating at a frequency of 2.50GHz and 16.0GB of RAM. The simulation software employed for these experiments is Matlab 2022b.
\subsection{IBWO performance on the 23 Standard Benchmark Functions}
To validate the performance of IBWO, 23 classical benchmark functions are applied \cite{suganthan2005problem}, as depicted in Table \ref{benchmark}. These benchmark functions are renowned for their strong typicality and are commonly employed for assessing the efficiency of optimization algorithms.

Optimization algorithms typically undergo evaluation using a set of benchmark functions that contain predefined optimal values. The efficacy of an algorithm is assessed based on the proximity of the results obtained from solving these benchmark problems to the predefined optimal values, as illustrated in Table \ref{benchmark}. Table \ref{IBWO_table} presents the outcomes derived from executing the optimization algorithm on these benchmark functions. Given the iterative nature of the algorithm, the table simultaneously presents both the average (\textbf{avg}) and standard deviation (\textbf{std}), serving as indicators of the algorithm's optimization performance and consistency, respectively. Alignment of the mean value with the predefined optimal value represents superior optimization performance. In cases where algorithms yield identical average results, the standard deviation is considered as a determinant of their stability. Generally, the effectiveness of optimization algorithms is measured by the degree of proximity between their average results and the known optimal values for each benchmark function.

In order to ensure equitable comparison of outcomes, we standardized the population size to 40 and restricted the iteration count to 200. Each algorithm was executed 30 times on each function independently to mitigate the influence of stochastic fluctuations on algorithmic performance.

\begin{figure*}[!t]
    \centering
    \subfloat[F21]{\includegraphics[width=1.65in]{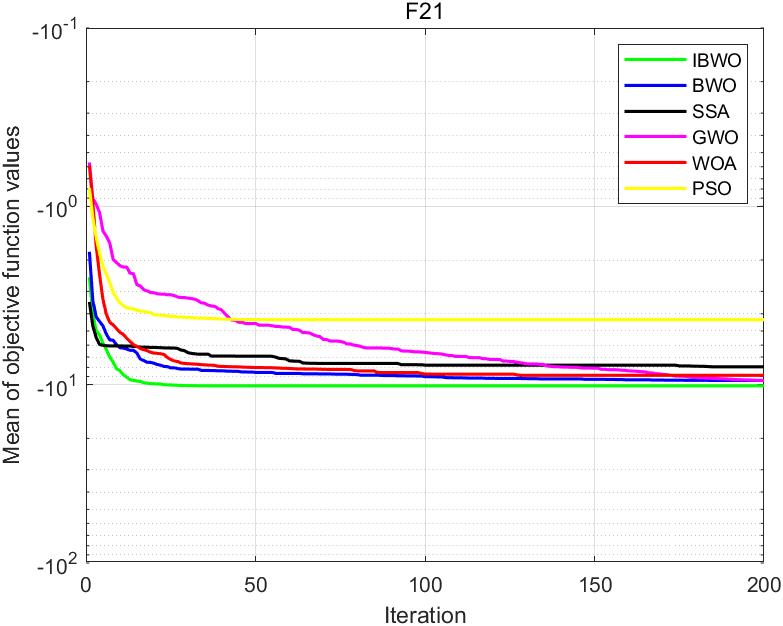}}%
    \label{figure21}
    \hfil
    \subfloat[F22]{\includegraphics[width=1.65in]{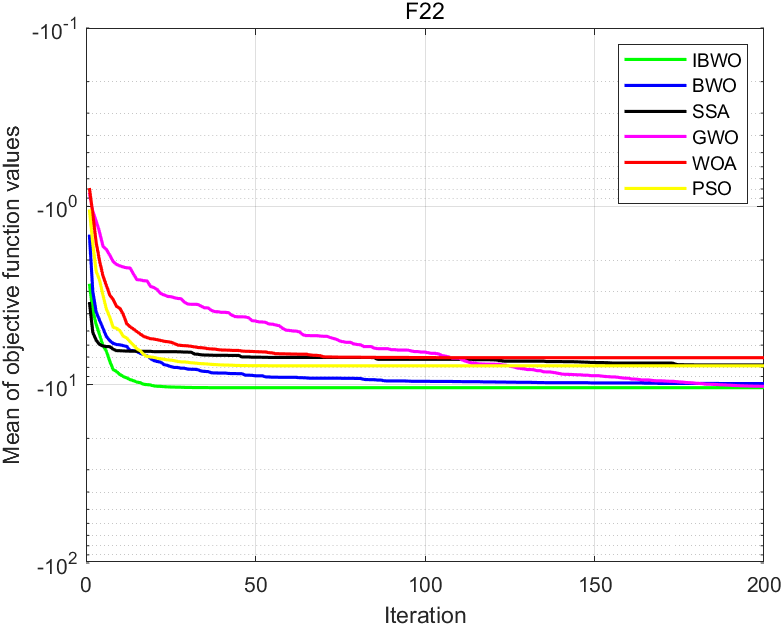}}%
    \label{figure22}
    \hfil
    \subfloat[F23]{\includegraphics[width=1.65in]{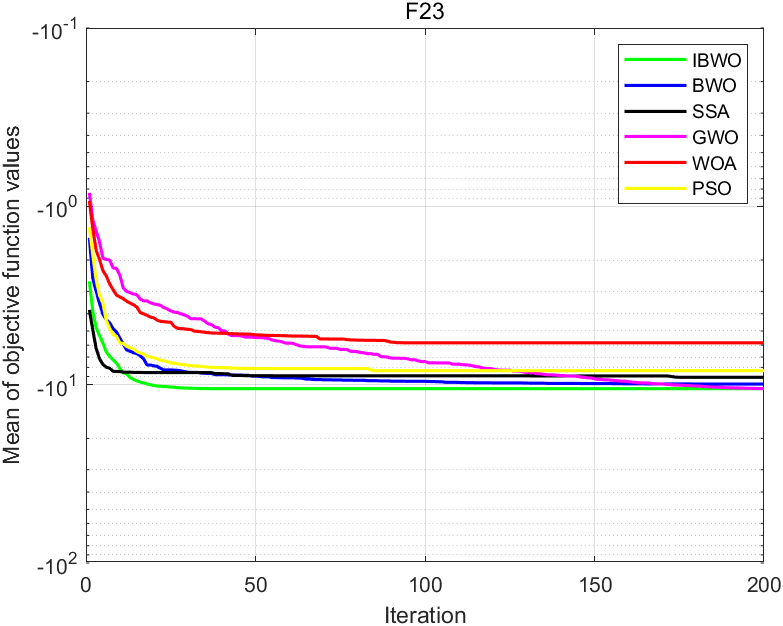}}%
    \label{figure23}\\
    \caption{The performance of IBWO and 5 other representative meta-heuristic algorithms on the F21-F23 benchmark functions.}
    \label{IBWO_graphics2}
\end{figure*}

We employ the average (\textbf{avg}) and standard deviation (\textbf{std}) derived from 30 distinct iterations as our evaluation criteria to evaluate the optimization efficacy of IBWO. These metrics were compared with the performance of the original Beluga Whale Optimization(BWO) \cite{zhong2022beluga}, Particle Swarm Optimization(PSO) \cite{marini2015particle}, Sparrow search algorithm(SSA) \cite{xue2020novel}, Grey Wolf Optimization(GWO) \cite{mirjalili2014grey}, and Whale Optimization algorithm(WOA) \cite{mirjalili2016whale} across 23 widely recognized benchmark functions. Furthermore, we depicted the trend of the algorithms' mean objective function values against the number of iterations, computed as an average over the results of 30 independent experiments. The results are graphically illustrated in Fig.\ref{IBWO_graphics1} and Fig.\ref{IBWO_graphics2}, and elaborated upon in Table \ref{IBWO_table}. The parameter configurations for these algorithms are outlined in Table \ref{PARAMETER}.

\begin{table}[htbp]
\caption{Parameters Configurations}
\centering
\begin{tabular}{@{}lll@{}}
\toprule
Algorithm & Parameters & Value \\ 
\midrule
WOA & Probability of encircling mechanism & 0.5 \\
   & Spiral factor & 1 \\
\addlinespace
PSO & \( c_1 \) & 2 \\
   & \( c_2 \) & 2 \\
   & \( V_{\text{max}} \) & 10 \\
\addlinespace
ISSA & \( ST \) & 0.8 \\
    & \( PD \) & 0.2 \\
    & \( SD \) & 0.1 \\
\addlinespace
GWO & \( a \) & [2,0] \\
BWO & Probability of whale fall decreased at
interval \( W_f \) & [0.1,0.05] \\
IBWO & Probability of whale fall decreased at
interval \( W_f \) & [0.1,0.05] \\
\bottomrule
\end{tabular}
\label{PARAMETER}
\end{table}

According to Table \ref{IBWO_table}, the optimization performance of IBWO across benchmark functions F1-F6, F9-F14, F16-19, and F21-F23 ranks first or ties for first among the six optimization algorithms. Furthermore, IBWO achieves the predefined optimal values on benchmark functions F9, F11, F16-F19, and F21-F23. As depicted in Fig.\ref{IBWO_graphics1} and \ref{IBWO_graphics2}, IBWO exhibits significantly faster convergence rates than the other five optimization algorithms on benchmark functions F1-F4, F9-F11, F14. Therefore, compared to the original BWO and other representative meta-heuristic algorithms, IBWO demonstrates outstanding performance in both optimization capability and convergence speed.

In summary, given IBWO's robust optimization capabilities and rapid convergence on most complex benchmark functions, it can be inferred that IBWO is adaptable to various types of non-convex objective functions. For the non-convex optimization problem of computing unknown robot positions in SRSs scenarios, IBWO emerges as a highly competitive candidate solution.

\subsection{Performance of the Proposed Localization Method}
\begin{figure}[htbp]
  \centering
  \begin{minipage}[b]{0.5\linewidth}  
    \centering
    \includegraphics[width=\linewidth]{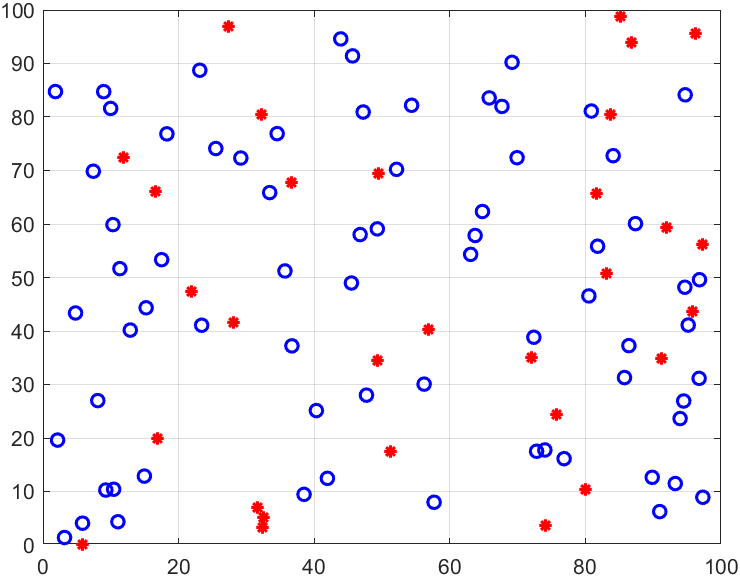} 
    \caption*{(a)Simulation environment} 
  \end{minipage}%
  \begin{minipage}[b]{0.5\linewidth}  
    \centering
    \includegraphics[width=\linewidth]{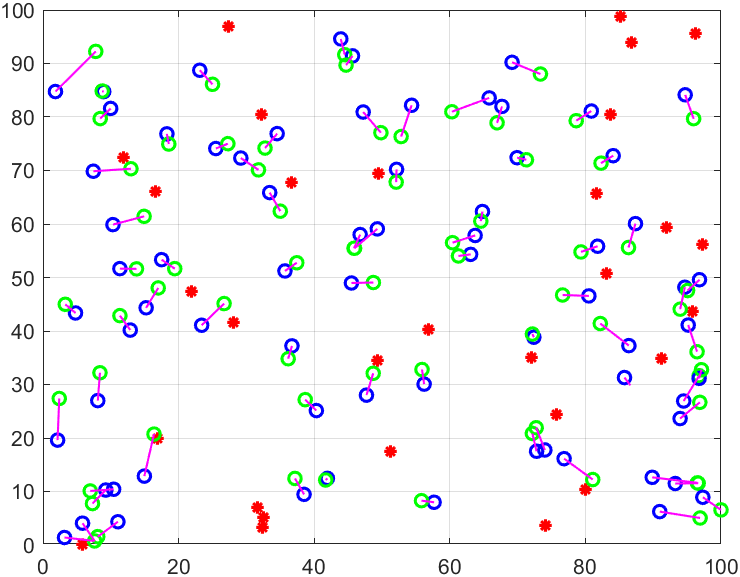} 
    \caption*{(b)Simulation result} 
  \end{minipage}
  \caption{The visual display of simulation environment and simulation result} 
  \label{visual}
\end{figure}

In this section, the proposed Improved Beluga Whale Localization (IBWOL) will be applied in a square obstacle-free simulated environment of size 100 by 100 meters. Fig.\ref{visual} (a) illustrates the simulation environment, the red nodes are the anchors and the blue nodes are the real position of the unknown robots. Fig.\ref{visual} (b) illustrates the localization results of the proposed positioning algorithm when the number of unknown robots is 70 and the number of anchors is 30. The green nodes represent the estimated positions of unknown robots, interconnected with the actual positions by purple lines. This visualization serves as a depiction, with a detailed analysis of the positioning algorithm provided in the subsequent sections. The evaluation criteria will include Normalized Relative Error (NRE) and Absolute Error (AE). Additionally, the performance of the algorithm will be assessed under three simulation scenarios: varying proportions of anchor, total number of robots, and communication radius of the robot. The traditional BWO algorithm will be employed within the proposed localization system for comparison, denoted as BWOL. Furthermore, other localization methods are employed for comparison include ISSAL \cite{zhang2021novel}, GWOL \cite{rajakumar2017gwo}, SAPSOL \cite{cheng2015self} and Multilateration \cite{wang2009localization}.

The computation method for Absolute Error(AE) and Normalized Relative Error(NRE) are defined in (25) and (26):
\begin{equation}
\text{AE} = \frac{\sum_{i=1}^{N} \sqrt{ (x_{\text{exact}}^i - x_{\text{estimated}}^i)^2 + (y_{\text{exact}}^i - y_{\text{estimated}}^i)^2 }}{N}
\end{equation}
\begin{equation}
\text{NRE} = \frac{\sum_{i=1}^{N} \sqrt{ (x_{\text{exact}}^i - x_{\text{estimated}}^i)^2 + (y_{\text{exact}}^i - y_{\text{estimated}}^i)^2 }}{N \times R}
\end{equation}

Where the coordinates\( (x_{\text{exact}}^{i},y_{\text{exact}}^{i}) \) represent the real position of the unknown robot, \( (x_{\text{estimated}}^{i},y_{\text{estimated}}^{i}) \) denote the estimated position of the unknown robot. N denotes the total number of unknown robots, and R represents the communication radius of robots.

\subsubsection{Varying Proportions of Anchor}
\begin{figure}[!htb]
  \centering
  \includegraphics[width=\linewidth]{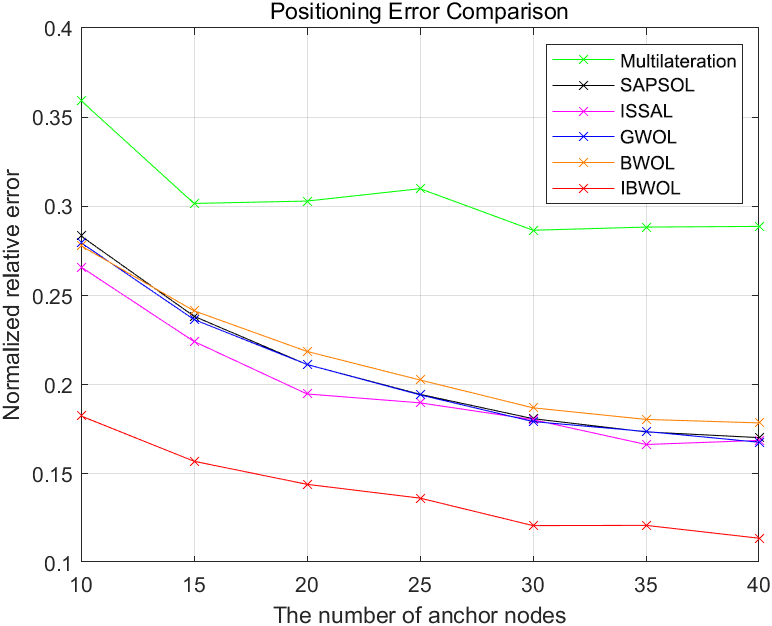}
  \caption{Normalized relative error vs. Proportions of  anchor}
  \label{anchor_proportions}
\end{figure}

In this experiment, we focus on comparing the performance of the proposed localization method with other localization methods and examining the impact of anchor node proportions on localization performance. The total node count and node communication radius are set to 100 and 30, respectively. The proportion of anchors is varied from 10 percent to 40 percent, with measurements taken at 5 percent intervals. The final results are presented in the form of localization error curves by NRE, as shown in Fig.\ref{anchor_proportions}. Additionally, the AE of this experiment is displayed in Table \ref{anchor_table}.

Based on Fig.\ref{anchor_proportions} and Table \ref{anchor_table}, it is evident that with the increase in the proportion of anchors, the localization error significantly decreases across all localization methods. This observation suggests a negative correlation between the proportion of anchors and both types of localization error under identical experimental conditions. Irrespective of the proportion of anchors, both the NRE and AE of the proposed localization method are consistently lower compared to the other five localization methods. In comparison to BWOL, ISSAL, SAPSOL, GWOL, and Multilateration, the proposed localization method demonstrates an average reduction in absolute error of 34.48\%, 29.86\%, 32.85\%, 32.39\%, and 54.38\% across all proportions of anchors. Additionally, when the proportion of anchors exceeds 20\%, the normalized relative error remains below 0.15, indicating highly competitive localization accuracy.

In conclusion, it can be inferred that the proposed localization method achieves favorable localization performance under limited resources of anchor proportions.

\begin{table}[htbp]
\caption{Absolute positioning error for different proportions of anchors}
\label{anchor_table}
\centering
\begin{tabular}{@{}lccccccc@{}}
\toprule
Anchor ratio & 10 & 15 & 20 & 25 & 30 & 35 & 40 \\ 
\midrule
IBWOL & 5.46 & 4.71 & 4.32 & 4.08 & 3.63 & 3.63 & 3.42 \\
BWOL & 8.34 & 7.23 & 6.57 & 6.09 & 5.61 & 5.43 & 5.37 \\
ISSAL & 7.98 & 6.72 & 5.85 & 5.70 & 5.40 & 4.98 & 5.07 \\
SAPSOL & 8.49 & 7.14 & 6.33 & 5.85 & 5.43 & 5.22 & 5.10 \\
GWOL & 8.37 & 7.11 & 6.33 & 5.82 & 5.37 & 5.22 & 5.04 \\
Multilateration & 10.77 & 9.06 & 9.09 & 9.30 & 8.61 & 8.64 & 8.64 \\
\bottomrule
\end{tabular}
\end{table}

\subsubsection{Varying the Communication Radius of Robots}
In this experiment, we mainly investigate the performance comparison of the proposed localization method with the other 5 localization methods under different communication radius scenarios, and examine the impact of communication radius on localization performance. The total number of robots and anchors is set to 100 and 30, respectively. The communication radius of robots is varied from 20m to 40m, with measurements taken at 5m intervals.

\begin{figure}[!htb]
  \centering
  \includegraphics[width=\linewidth]{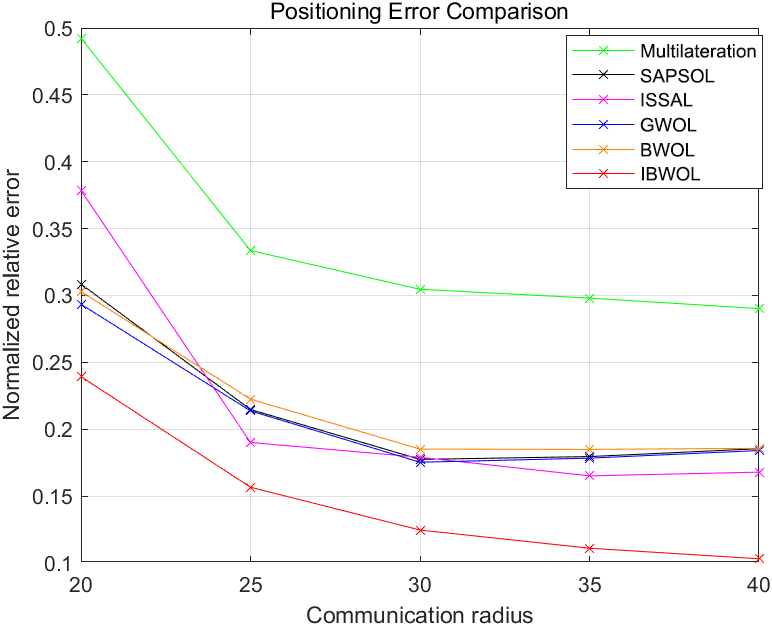}
  \caption{Normalized relative error vs. Communication radius}
  \label{radius_figure}
\end{figure}

The results from Fig.\ref{radius_figure} and Table \ref{radius_table} indicate that with the increase in the communication radius \(R\), the AE of all localization algorithms initially decreases and then increases. The decrease occurs because at a smaller communication radius, the minimum number of hops between nodes is larger, and increasing the communication radius can reduce the minimum number of hops between nodes. Thereby reducing the cumulative error during distance measurement. However, the increase occurs because as the communication radius increases, especially when it exceeds 30m, more and more pairs of nodes have a minimum hop count of 1, regardless of whether their real distances are close or far. This significantly affects the calculation of the average hop distance, thereby impacting the localization accuracy. In contrast, in NRE, the presence of the communication radius \( R \) in the denominator of its calculation formula offsets the effect of the increased communication radius, resulting in an overall downward trend. Additionally, both AE and NRE of the proposed localization algorithm are consistently lower than the other 5 localization algorithms across all communication radius. Compared to BWOL, ISSAL, SAPSOL, GWOL, and Multilateration, The proposed algorithm exhibits a reduction in average absolute error of 32.69\%, 30.49\%, 31.24\%, 30.28\%, and 58.05\%, respectively. Thus, it can be inferred that the proposed algorithm demonstrates superior and stable performance across various communication radius.

\begin{table}[htbp]
\caption{Absolute positioning error for different communication radius}
\label{radius_table}
\centering
\begin{tabular}{@{}lccccccc@{}}
\toprule
Communication radius & 20 & 25 & 30 & 35 & 40 \\ 
\midrule
IBWOL & 5.10 & 4.16 & 3.70 & 3.80 & 4.08  \\
BWOL & 6.03 & 5.76 & 5.69 & 5.91 & 7.57  \\
ISSAL & 6.62 & 5.65 & 5.24 & 5.58 & 6.89 \\
SAPSOL & 5.81 & 5.53 & 5.56 & 5.94 & 7.47  \\
GWOL & 5.79 & 5.34 & 5.47 & 5.86 & 7.43 \\
Multilateration & 11.24 & 8.43 & 8.42 & 10.24 & 11.35  \\
\bottomrule
\end{tabular}
\end{table}

\subsubsection{Varying the total number of nodes}
\begin{figure}[!htb]
  \centering
  \includegraphics[width=\linewidth]{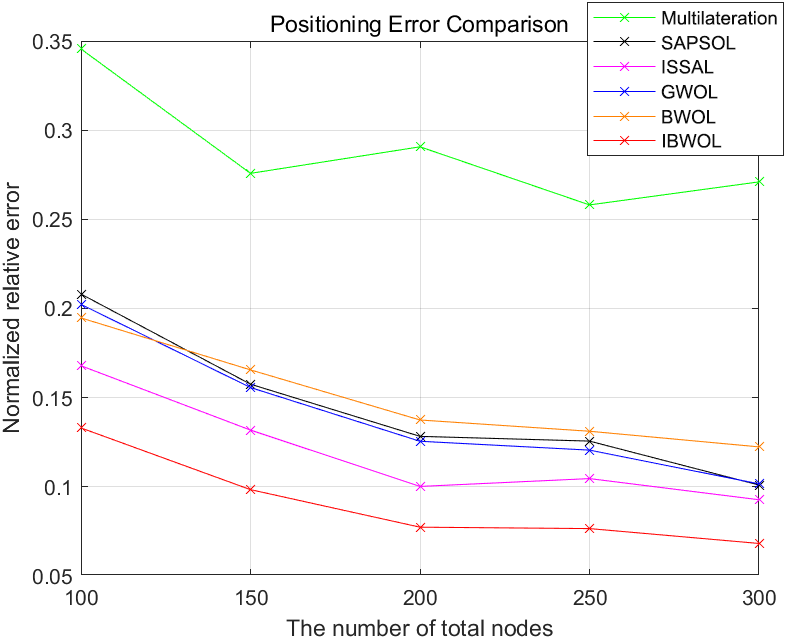}
  \caption{Normalized relative error vs. The total number of  robots}
  \label{total_figure}
\end{figure}

In this experiment, we primarily study the performance comparison of the localization algorithm proposed in this study with 5 other positioning algorithms under varying total number of robots, as well as the impact of the total number of robots on the performance of the localization algorithm. The total number of robots is incrementally increased from 100 to 300, with the positioning error measured at intervals of 50 robots. The proportion of anchors is set at 30\%, and the communication radius is set to 30m. The NRE of the positioning algorithm is depicted in Fig.\ref{total_figure}, while the AE can be found in Table \ref{total_table}.

Based on Fig.\ref{total_figure} and Table \ref{total_table}, the two types of errors for all six positioning methods decrease as the total number of robots increases, suggesting a negative correlation between the total number of robots and positioning error. Moreover, regardless of the total number of robots, the two types of positioning errors of the method proposed in this paper are consistently lower than those of the other five positioning methods. Compared to BWOL, ISSAL, SAPSOL, GWOL, and Multilateration, the proposed absolute localization errors have been reduced by 39.68\%, 24.25\%, 37.17\%, 35.93\%, and 68.59\%, respectively. Furthermore, when the total number of robots exceeds 150, the NRE of IBWOL is below 0.1, which achieves a remarkably low level. In summary, the proposed localization algorithm demonstrates superior localization accuracy across different numbers of nodes.

\begin{table}[htbp]
\caption{Absolute positioning error for different total number of robots}
\label{total_table}
\centering
\begin{tabular}{@{}lccccccc@{}}
\toprule
Number of total nodes & 100 & 150 & 200 & 250 & 300 \\ 
\midrule
IBWOL & 3.99 & 2.94 & 2.31 & 2.31 & 2.04 \\
BWOL & 5.85 & 4.98 & 4.11 & 3.93 & 3.66  \\
ISSAL & 5.04 & 3.96 & 3.00 & 3.15 & 2.79 \\
SAPSOL & 6.24 & 4.74 & 3.84 & 3.78 & 3.03 \\
GWOL & 6.06 & 4.68 & 3.78 & 3.63 & 3.06  \\
Multilateration & 10.38 & 8.28 & 8.73 & 7.74 & 8.13 \\
\bottomrule
\end{tabular}
\end{table}

\section{Conclusion}
This study proposes a novel Improved Beluga Whale Optimization (IBWO) algorithm to solve the localization problem in swarm robotics. To enhance information sharing among the population, we incorporated CSFS and CCFS during the exploitation phase of the BWO. To improve the algorithm's ability to escape local optimal, we add the Golden-SA following the whale fall phase of the BWO. Simulations on 23 classical test functions demonstrate that the IBWO exhibits superior optimization capabilities compared to the BWO and four other meta-heuristic algorithms.

For the swarm robotics localization problem, this study first uses Hop-optimized DV-Hop to measure the distances between unknown robots and anchors. The localization problem is then abstracted as a non-convex optimization problem, and an objective function is constructed using the distances between the unknown robots and anchors. Finally, the Improved Beluga Whale Optimization (IBWO) algorithm is used to optimize this objective function. Tests in 100x100 simulation area demonstrate that the proposed method achieves higher localization accuracy compared to traditional Multilateration and four other localization methods.

As future work, to further improve the localization accuracy of robots, we will continue to explore meta-heuristic algorithms with stronger optimization capabilities. Additionally, we plan to modify objective functions by using the position relationships between unknown robots. In real-world scenarios, the presence of obstacles can block signal transmission between robots, leading to significant distance measurement errors. So ensuring the localization accuracy of robots in the presence of obstacles will also be a direction of our research.


%

\ifCLASSOPTIONcaptionsoff
  \newpage
\fi

\bibliographystyle{IEEEtran}
\bibliography{ref}

\end{document}